\documentclass[aps,prd,amsmath,preprint,floatfix]{revtex4}

\usepackage[utf8]{inputenc}
\usepackage{enumerate}
\usepackage{amsmath,amssymb}
\usepackage{bm}
\usepackage{amscd}      
\usepackage{graphicx}
\usepackage{hhline,multirow,tabularx}  
\usepackage{ulem}  	
\usepackage{dcolumn}    
\usepackage{url}        
\usepackage{color}
\usepackage[dvipsnames]{xcolor}

\preprint{JLAB-THY-16-2215}

\begin{document}

\title{Constraints on large-$x$ parton distributions from new \\
	weak boson production and deep-inelastic scattering data}

\author{A.~Accardi$^{1,2}$,
	L.~T.~Brady$^{2,3}$,
	W.~Melnitchouk$^2$,
	J.~F.~Owens$^4$,
	N.~Sato$^2$}

\affiliation{
$^1$\mbox{Hampton University, Hampton, Virginia 23668, USA} \\
$^2$\mbox{Jefferson Lab, Newport News, Virginia 23606, USA} \\
$^3$\mbox{University of California, Santa Barbara, California 93106, USA} \\
$^4$\mbox{Florida State University, Tallahassee, Florida 32306, USA}}

\date{\today}

\begin{abstract}
We present a new set of leading twist parton distribution functions,
referred to as ``CJ15'', which take advantage of developments in the
theoretical treatment of nuclear corrections as well as new data.
The analysis includes for the first time data on the free neutron
structure function from Jefferson Lab, and new high-precision
charged lepton and $W$-boson asymmetry data from Fermilab.
These significantly reduce the uncertainty on the $d/u$ ratio at
large values of $x$ and provide new insights into the partonic
structure of bound nucleons.
\end{abstract}

\maketitle

\section{Introduction}
\label{sec:intro}

Tremendous advances have been made over the past decade in our
knowledge of the quark and gluon (or parton) substructure of the
nucleon, with the availability of new high energy scattering data
from various accelerator facilities worldwide
\cite{JMO13, Blumlein13, ForteWatt13}.
Results from the final analysis of data from the $ep$ collider
HERA have allowed a detailed mapping of the partonic landscape
at small values of the nucleon's parton momentum fraction $x$
\cite{Cooper-Sarkar15}.
Data from high energy $p\bar p$ scattering at the Tevatron on weak
boson and jet production have provided a wealth of complementary
information on the nucleon's flavor structure.
At lower energies, precision structure function measurements at the
high luminosity CEBAF accelerator at Jefferson Lab have enabled a
detailed investigation of nucleon structure at large values of $x$
\cite{JLab11}.
More recently, fascinating glimpses into the role played by sea
quarks and gluons in the proton have been seen in various channels
in $pp$ collisions at the LHC.

To analyze the vast amounts of data from the various facilities,
concerted efforts are being made to systematically extract information
about the nucleon's quark and gluon structure in the form of parton
distribution functions (PDFs) \cite{MMHT14, CT14, NNPDF3.0, HERAPDF15,
JR14, ABMP15, CJ10, CJ11, CJ12}.  While much of the effort has in the
past been directed at the small-$x$ frontier made accessible through the
highest energy colliders, relatively less attention has been focused
on the region of large momentum fractions, where nonperturbative QCD
effects generally play a more important role.

The CTEQ-Jefferson Lab (CJ) collaboration \cite{CJweb} has performed
a series of global PDF analyses \cite{CJ10, CJ11, CJ12} with the aim
of maximally utilizing data at the highest $x$ values amenable to
perturbative QCD treatment.  The additional complications of working
with data down to relatively low values of four-momentum transfer
$Q^2$ ($Q^2 \gtrsim 1-2$~GeV$^2$) and invariant final state masses
$W^2$ ($W^2 \gtrsim 4$~GeV$^2$) have been met with developments in
the theoretical description of various effects which come into
prominence at such kinematics.
The importance of $1/Q^2$ power corrections, arising from target
mass and higher twist effects, has been emphasized \cite{CJ10, CJ11}
particularly in the analysis of fixed-target deep-inelastic scattering
(DIS) data, which found leading twist PDFs to be stable down to low
$Q^2$ values with the inclusion of both of these effects.

Moreover, since the CJ analyses typically fit both proton and deuterium
data, the description of the latter requires careful treatment of
nuclear corrections at large values of $x$, at all $Q^2$ scales.
The $d$-quark PDF is especially sensitive to the deuterium corrections
for $x \gtrsim 0.5$, and historically has suffered from large
uncertainties due to the model dependence of the nuclear effects
\cite{MT96}.
To adequately allow for the full range of nuclear model uncertainties,
the CJ12 analysis \cite{CJ12} produced three sets of PDFs corresponding
to different strengths (minimum, medium and maximum) of the nuclear
effects, which served to provide a more realistic estimate of the
$d$-quark PDF uncertainty compared with previous fits.

In this analysis, which we refer to as ``CJ15'', we examine the impact
of new large rapidity charged lepton and $W$-boson asymmetry data
from the Tevatron \cite{D0_mu, D0_e, D0_W} on the determination of
next-to-leading order (NLO) PDFs and their errors, particularly at
large values of $x$.
We also include for the first time new Jefferson Lab data on the
free neutron structure function obtained from backward spectator
proton tagging in semi-inclusive DIS \cite{Baillie12, Tkachenko14},
which do not suffer from the same uncertainties that have afflicted
previous neutron extractions.
We present a more complete treatment of the nuclear corrections
in deuterium, examining a range of high-precision deuteron wave
functions and several models for the nucleon off-shell corrections.
In contrast to our earlier fits \cite{CJ10, CJ11, CJ12}, which
relied on physically-motivated models for the off-shell effects,
the precision of the new data allows us to perform a purely 
phenomenological fit, with the off-shell parameters determined
directly from the data.
Other improvements in the CJ15 analysis include a more robust
parametrization of the $\bar d/\bar u$ asymmetry, which accommodates
different asymptotic behaviors as $x \to 1$, and the implementation
of the S-ACOT scheme \cite{S-ACOT} for heavy quarks.

In Sec.~\ref{sec:thy} we review the theoretical formalism underpinning
our global analysis, including the choice of parametrization for the
various PDFs.  We discuss the treatment of mass thresholds, and the
application of finite-$Q^2$ corrections from target mass and higher
twist effects that are necessary to describe the low-$Q^2$,
large-$x$ data.
A detailed investigation of nuclear corrections in the deuteron
follows, in which we outline several models and parametrizations of
nucleon off-shell corrections, which represent the main uncertainty
in the computation of the nuclear effects.

In Sec.~\ref{sec:data} a summary of the data sets used in this
analysis is given, and the results of the fits are presented in
Sec.~\ref{sec:results}.  Here we compare the CJ15 PDFs with other
modern parametrizations, as well as with selected observables.
In addition to the NLO analysis, we also perform a leading order
(LO) fit, which is useful for certain applications, such as
Monte Carlo generators or for estimating cross sections and
event rates for new experiments.
Our central results deal with the role played by the nuclear
corrections and their uncertainties in the global analysis,
and how these can be reduced by exploiting the interplay of
different observables sensitive to the $d$-quark PDF.
We discuss the consequences of the new analysis for the shape
of the deuteron to isoscalar nucleon structure function ratio,
and the closely related question of the behavior of the $d/u$
PDF ratio at large $x$.
Finally, in Sec.~\ref{sec:conclusion} we summarize our results and
discuss possible future improvements in PDF determination that
are expected to come with new data from collider and fixed-target
experiments.

\section{Theoretical foundations}
\label{sec:thy}

In this section we present the theoretical framework on which the
CJ15 PDF analysis is based.  We begin with a discussion of the
parametrizations chosen for the various flavor PDFs, noting the
particular forms used here for the $d/u$ and $\bar d/\bar u$ ratios
compared with earlier work.  We then describe our treatment of
heavy quarks, and the implementation of finite-$Q^2$ corrections.
A detailed discussion of the nuclear corrections in the deuteron
follows, where we review previous attempts to account for nucleon
off-shell effects, and describe the approach taken in this analysis.

\subsection{PDF parametrizations}
\label{ssec:parametrizations}

For the parametrization of the PDFs at the input scale $Q_0^2$,
chosen here to be the mass of the charm quark, $Q_0^2 = m_c^2$,
a standard 5-parameter form is adopted for most parton species~$f$,
\begin{equation}
xf(x,Q_0^2) = a_0\, x^{a_1} (1-x)^{a_2}\, (1 + a_3 \sqrt{x} + a_4 x).
\label{eq:param}
\end{equation}   
This form applies to the valence $u_v = u-\bar u$ and $d_v = d-\bar d$
distributions, the light antiquark sea $\bar u + \bar d$, and the
gluon distribution $g$.  The charm quark is considered to be
radiatively generated from the gluons.
To allow greater flexibility for the valence $d_v$ PDF in the
large-$x$ region, we add in a small admixture of the valence
$u$-quark PDF,
\begin{equation}
d_v \rightarrow
    a_0^{d_v} \left( \frac{d_v}{a_0^{d_v}} + b\, x^c\, u_v \right),
\label{eq:du}
\end{equation}
with $b$ and $c$ as two additional parameters.
The result of this modification is that the ratio
	$d_v/u_v \to a_0^{d_v}\, b$ as $x \to 1$,
provided that $a_2^{d_v} > a_2^{u_v}$, which is usually the case.
This form avoids potentially large biases on the $d$-quark PDF
central value \cite{CJ11}, as well as on its PDF error estimate
\cite{Accardi13}, as we discuss in detail in Sec.~\ref{sec:results}.
A finite, nonzero value of the $d_v/u_v$ ratio is also expected
in several nonperturbative models of hadron structure
\cite{MT96, FJ75, Close73, Close03, HR10, Roberts13}.
The normalization parameters $a_0$ for the $u_v$ and $d_v$
distributions are fixed by the appropriate valence quark number
sum rules, while $a_0^g$ is fixed by the momentum sum rule.

In the CJ12 PDF sets the combinations $\bar d \pm \bar u$ were
parametrized separately.  In that analysis it was found to be
difficult to control the size of the $\bar d$ distribution
relative to the $\bar u$ at values of $x$ above about 0.3,
since there were essentially no constraints on the sea quarks.
Consequently some fits generated $\bar d$ PDFs that became negative
in this region.  While this had little effect on the NLO fits since
the terms were very small there, it was nonetheless unsatisfactory
when one considered LO fits where the PDFs are expected to be
positive.
In the present analysis we therefore parametrize directly the
ratio $\bar d/\bar u$ instead of the difference $\bar d - \bar u$.
For the functional form of $\bar d/\bar u$ at the input scale
$Q_0^2$ we choose 
\begin{eqnarray}
\frac{\bar d}{\bar u}
&=& a_0 x^{a_1} (1-x)^{a_2} + 1 + a_3 x (1-x)^{a_4},
\label{eq:dboub}
\end{eqnarray}
which ensures that in the limit $x \to 1$ one has $\bar d/\bar u \to 1$.
This is actually a theoretical prejudice since the sea quark PDFs are
fed by $Q^2$ evolution which, in the absence of isospin symmetry
violating effects, generates equal $\bar d$ and $\bar u$ contributions.

Since the existing data are not able to reliably determine the
large-$x$ behavior of the ratio, we have also performed alternative
fits using
  $\bar d/\bar u = a_0 x^{a_1} (1-x)^{a_2} + (1 + a_3 x) (1-x)^{a_4}$,
which vanishes in the $x \to 1$ limit.
Data from the E866 dilepton production experiment \cite{E866, E866rat}
currently provide the strongest constraints on the $\bar d/\bar u$
ratio and show a decrease below unity in the region $x\gtrsim 0.3$,
albeit with large errors.
It was found that either parametrization could achieve excellent fits
to the data included in the global analysis.
In the central fits presented here we choose the parametrization
in Eq.(\ref{eq:dboub}).  In the near future the SeaQuest experiment
(E906) \cite{SeaQuest} at Fermilab is expected to yield data with
higher statistical precision that will constrain the $\bar d/\bar u$
ratio to larger $x$ values, and so answer the question as to which
parametrization is more suitable.
Data on Drell-Yan and $W$-boson production in $pp$ collisions at the
LHC should also provide important constraints on the behavior of
$\bar d/\bar u$ outside of currently accessible regions of $x$.

For the strange quark distribution, the strongest constraints have
traditionally come from charm meson production in neutrino DIS from
nuclei.  In keeping with the approach adopted in our previous analyses
\cite{CJ10, CJ11, CJ12}, we do not include neutrino scattering data
in the current fit because of uncertainties in relating structure
functions of heavy nuclei to those of free nucleons.
Moreover, a proper treatment of dimuon production on nuclear targets
requires inclusion of initial state as well as final state nuclear
effects \cite{Accardi09, Majumder11}.
The former are relatively well understood and accounted for by
using nuclear PDFs \cite{Eskola09, deFlorian11, Kovarik15}.
The latter, however, include effects such as the scattered charm quark
energy loss while traversing the target nucleus, or $D$ meson--nucleon
interactions for mesons hadronizing within the nucleus, which are much
less known.
These effects have often been underestimated theoretically in the
analysis of heavy-ion reactions, and are essentially unknown
experimentally in nuclear DIS, constituting a potentially large
source of systematic uncertainty.
Consequently, we follow our previous strategy in assuming flavor
independence of the shape of the sea quark PDFs, with a fixed ratio
\begin{eqnarray}
\kappa
&=& \frac{s + \bar s}{\bar u + \bar d}.
\label{eq:kappa}
\end{eqnarray}
We further make the usual assumption that $\bar s = s$,
and take $\kappa = 0.4$.

Recently an analysis of ATLAS data on $W$ and $Z$ production in $pp$
collisions at the LHC claimed a significantly larger strange quark
sea, with $\kappa \sim 1$ \cite{ATLAS-s}.
However, in a combined fit to data on charm production from neutrino
DIS and from the LHC, Alekhin {\it et al.} \cite{Alekhin-s} argued
that the apparent strange quark enhancement is likely due to a
corresponding suppression of the $\bar d$-quark PDF at small $x$.
They point out that this reflects the limitations of attempts to
separate individual quark flavor PDFs based solely on data from
$pp {\rm \ and \ }ep$ scattering. 
Note that in all of these analyses the assumption is made that
$s = \bar s$.  Possible differences between the $s$ and $\bar s$
PDFs can arise from both perturbative \cite{Catani04} and
nonperturbative \cite{Signal87} effects and could affect,
for example, the extraction of the weak mixing angle from
neutrino data \cite{Zeller02}.
A detailed analysis of the strange quark PDF using LHC and other
data within the CJ framework will be performed in future work.

\subsection{Heavy quarks}
\label{ssec:HQs}

The existence of heavy quark PDFs in the nucleon introduces new mass
scales and leads to the appearance of logarithmic terms of the form
$\log Q^2/m_q^2$ in perturbative QCD calculations, where $m_q$ is
the mass of the heavy quark.
As $Q^2$ grows these can become large and need to be resummed.
The evolution equations for the PDFs sum these potentially large
logarithms.  In schemes where the heavy quarks are treated as massless
in the hard scattering subprocesses, the heavy quark mass enters via
the boundary conditions on the PDFs at the heavy quark threshold.
Typically this takes the form of imposing that the heavy quark PDF
vanishes for $Q^2$ below $m_q^2$, with massless evolution being used
as $Q^2$ increases.  Although valid asymptotically, this result does
not treat the threshold region correctly, since the threshold occurs
in the variable $W^2$, not $Q^2$.

In this analysis we are interested in determining the PDFs over ranges
of $Q^2$ and $x$ that include the threshold regions for the $c$ and
$b$ quarks.  To correctly treat these regions we implement the S-ACOT
scheme as presented in Ref.~\cite{S-ACOT}.  This is a simplified
version that is equivalent to the variable flavor ACOT scheme
\cite{ACOT}.  The S-ACOT scheme has been implemented for the neutral
current DIS processes in the present analysis and we take the masses
of the charm and bottom quarks to be $m_c = 1.3$~GeV and
$m_b = 4.5$~GeV, respectively.

\subsection{$1/Q^2$ corrections}
\label{ssec:Q2}

The cuts on $Q^2$ and $W^2$ imposed on the data sets used in this
analysis (see Sec.~\ref{sec:data} below) significantly increase
the number of data points available to constrain PDFs.  While this
allows access to a greater range of kinematics and leads to reduced
PDF uncertainties, especially at higher values of $x$, it also
requires careful treatment of subleading, ${\cal O}(1/Q^2)$ power
corrections to the leading twist calculations.
The most basic correction arises from imposing exact kinematics on
twist-two matrix elements at finite values of $Q^2$, which gives
rise to effects that scale with $x^2 M^2/Q^2$.  These target mass
corrections (TMCs) were first evaluated within the operator product
expansion (OPE) \cite{GP76} for DIS processes, and allow structure
functions at finite $Q^2$ to be expressed in terms of their 
$M^2/Q^2 \to 0$ (or ``massless'') values.  For the $F_2$ structure
function, for instance, one has \cite{GP76, Schienbein08, Brady11}
\begin{eqnarray}
F_2(x,Q^2)
&=& {(1+\rho)^2 \over 4 \rho^3} F_2^{(0)}(\xi,Q^2)\
 +\ {3x(\rho^2-1) \over 2\rho^4}
    \int_\xi^1 du\,
    \left[ 1 + {\rho^2-1 \over 2 x \rho} (u-\xi) \right]
    {F_2^{(0)}(u,Q^2) \over u^2},	\nonumber\\
& &
\label{eq:OPE_F2}
\end{eqnarray}
where $F_2^{(0)}$ is the structure function in the $M^2/Q^2 \to 0$
limit.  Here the massless limit functions are evaluated in terms of
the modified scaling variable $\xi$ \cite{Greenberg71, Nachtmann73},
\begin{equation}
\xi = {2x \over 1 + \rho}\, ,\ \ \ \ \ \ \ 
\rho^2 = 1 + \frac{4 x^2 M^2}{Q^2},
\end{equation}
which approaches $x$ as $M^2/Q^2 \to 0$.

Later work within the collinear factorization framework provided an
alternative formulation of TMCs \cite{EFP76}, which had the advantage
that it could also be applied to processes other than inclusive DIS
\cite{AHM09, Guerrero15}.
To ${\cal O}(1/Q^2)$ the two approaches can in fact be shown to be
equivalent.  A number of other prescriptions have also been proposed
in the literature \cite{Schienbein08, AOT94, KR02, AQ08, SBMS12},
using different approximations to the OPE and collinear factorization
methods.
In the context of a global PDF fit, it was found in Ref.~\cite{CJ10}
that differences arising from the various prescriptions can be
effectively compensated by the presence of phenomenological higher
twist terms.
In the present analysis we use the standard OPE expression for
the TMCs in Eq.~(\ref{eq:OPE_F2}).

For other subleading $1/Q^2$ corrections, which include higher twists
but also other residual power corrections, we follow our earlier work
\cite{CJ10, CJ11, CJ12} and parametrize the correction in terms of a
phenomenological $x$-dependent function,
\begin{align}
F_2(x,Q^2)
= F_2^{\rm LT}(x,Q^2)
  \left( 1 + \frac{C_{\rm HT}(x)}{Q^2} \right),
\label{eq:F2ht}
\end{align}
where $F_2^{\rm LT}$ denotes the leading twist structure function,
including TMCs.  The higher twist coefficient function is parametrized
by a polynomial function as
\begin{align}
C_{\rm HT}(x) = h_0\, x^{h_1} (1+h_2\, x),
\label{eq:C_ht}
\end{align}
with $h_0, h_1$ and $h_2$ as free parameters.
For simplicity we assume the high twist correction to be isospin
independent (see, however, Refs.~\cite{Vir92, AKL03, BB08, Blu12});
the possible isospin dependence of $F_2$ and other structure functions
will be studied in a future dedicated analysis.

\subsection{Nuclear corrections}
\label{ssec:nuclear}

As in the previous CJ PDF analyses \cite{CJ10, CJ11, CJ12}, the use of
deuterium DIS and Drell-Yan data necessitates taking into account the
differences between PDFs in the deuteron and those in the free proton
and neutron.  The CJ15 analysis follows a similar approach, with several
improvements over the earlier implementations, as we discuss in this
section.  While the earlier analyses applied nuclear corrections only
to deep-inelastic deuteron structure functions, here we formulate
the corrections at the parton level and generalize the treatment to
any process involving quark, antiquark or gluon PDFs in the deuteron.

Generally, the nuclear corrections in high energy reactions account
for Fermi motion, binding, and nucleon off-shell effects, which are
implemented in the form of convolutions with nuclear smearing
functions.  The nuclear effects become increasingly important at
intermediate and large values of $x$, and will be the focus of this
section.  In addition, rescattering effects mediated by Pomeron and
pion exchange mechanisms give rise to shadowing at small values of
$x$ ($x \lesssim 0.1$) \cite{Badelek92, MTshad} and a small amount
of antishadowing at $x \sim 0.1$ \cite{MTshad, Kaptari91}; in this
analysis we implement these using the results from Ref.~\cite{MTshad}.
In practice, however, the shadowing and antishadowing corrections
are very small, and have negligible effect on the phenomenology 
considered in this paper.

\subsubsection{Nuclear smearing}
\label{sssec:smear}

From the standard nuclear impulse approximation for the scattering
of a projectile (lepton or hadron) from a deuteron $d$, the
momentum distribution of a parton inside the deuteron is given
by a convolution of the corresponding PDF in the bound nucleon
and a momentum distribution $f_{N/d}$ of nucleons in the
deuteron (or ``smearing function'').
Taking for illustration the PDF for a quark of flavor $q$ (the
generalization to antiquarks and gluons is straightforward),
its parton momentum distribution in the deuteron can be computed
as \cite{MSToff, KPW94}
\begin{eqnarray}
q^d(x,Q^2)
&=& \int \frac{dz}{z} dp^2\,
    f_{N/d}(z,p^2)\, \widetilde{q}^N(x/z,p^2,Q^2),
\label{eq:genconv}
\end{eqnarray}
where $z = (M_d/M)(p \cdot q / p_d \cdot q)$ is the nucleon momentum
fraction in the deuteron, with $p$ and $p_d$ the four-momenta of the
nucleon and deuteron, respectively, and $M_d$ is the deuteron mass.
The nucleon virtuality $p^2$ defines the degree to which the bound
nucleon is off its mass shell, $p^2 \neq M^2$, and the function
$\widetilde{q}^N$ represents the quark PDF in the off-shell nucleon.
For the isoscalar deuteron, a sum over the nucleons $N = p, n$ is
implied.
For inclusive DIS, the higher twist contribution to the $F_2^d$
structure function is computed analogously to Eq.~(\ref{eq:genconv}),
by convoluting the product of the off-shell nucleon parton
distribution and the higher twist function $C_{\rm HT}$ in
Eq.~(\ref{eq:C_ht}) with the nucleon smearing function.
%
%
%
For DY processes the higher twist component is very small and can be
neglected \cite{Ehlers14}.

While the off-shell nucleon PDF $\widetilde{q}^N$ is not by itself an
observable, its dependence on the virtuality $p^2$ can be studied within
a given theoretical framework.  Since the bound state effects in the
deuteron are the smallest of all the atomic nuclei, one may expand the
off-shell nucleon distribution to lowest order about its on-shell limit
\cite{KPW94, KMPW95},
\begin{eqnarray}
\widetilde{q}^N (x,p^2,Q^2)
&=& q^N(x,Q^2)
    \left( 1 + \frac{p^2-M^2}{M^2} \delta f^N(x,Q^2) \right),
\label{eq:qoff}
\end{eqnarray}     
where the coefficient of the off-shell term is given by
\begin{eqnarray}
\delta f^N(x,Q^2)
&=& \left.
    \frac{\partial \ln \widetilde{q}^N(x,p^2,Q^2)}
	 {\partial \ln p^2}
    \right|_{p^2=M^2}.
\label{eq:deltafN}
\end{eqnarray}
The on-shell term in Eq.~(\ref{eq:qoff}) leads to the standard
on-shell convolution representation for the nuclear PDF, while
the off-shell term can be evaluated as an additive correction.
Defining the total quark PDF in the deuteron as the sum of the
on-shell and off-shell contributions,
  $q^d = q^{d\, (\rm on)} + q^{d\, (\rm off)}$,
the two components can be written as
\begin{subequations}
\label{eq:qd}
\begin{eqnarray}
q^{d\, (\rm on)}(x,Q^2)
&=& \int \frac{dz}{z} f^{(\rm on)}(z)\, q^N(x/z,Q^2),
\label{eq:qdon}						\\
q^{d\, (\rm off)}(x,Q^2)
&=& \int \frac{dz}{z} f^{(\rm off)}(z)\,
		      \delta f^N(x/z,Q^2)\, q^N(x/z,Q^2).
\label{eq:qdoff}
\end{eqnarray}  
\end{subequations}
The on-shell and off-shell smearing functions $f^{(\rm on)}$
and $f^{(\rm off)}$ are taken to be the same for the proton
and neutron (isospin symmetry breaking effects are not expected
to be significant) and are given by \cite{Ehlers14}
\begin{subequations}
\begin{eqnarray}
f^{(\rm on)}(z)
&=& \int dp^2\, f_{N/d}(z,p^2),		\\
f^{(\rm off)}(z)
&=& \int dp^2\, \frac{p^2-M^2}{M^2}\, f_{N/d}(z,p^2).
\end{eqnarray}
\end{subequations}
A systematic method for computing the smearing functions is within
the weak binding approximation, in terms of the deuteron wave
function \cite{KP06, KMK09}. For large $Q^2 \to \infty$ the on-shell
smearing function $f^{(\rm on)}$ has a simple probabilistic
interpretation in terms of the light-cone momentum fraction
  $z \to (M_d/M)(p^+/p_d^+)$
of the deuteron carried by the struck nucleon, where
$p^+ = p_0 + p_z$ is the ``plus'' component of the four-vector $p$.
At finite $Q^2$, however, the smearing functions depend also
on the parameter $\rho^2$, which characterizes the deviation
from the Bjorken limit, and the momentum fraction variable is
  $z = 1 + (\varepsilon + \rho\, p_z)/M)$,
where $\varepsilon = p_0 - M$ is the separation energy.
In fact, the $\rho$ dependence of the smearing functions is
different for the $F_1$ and $F_2$ DIS structure functions,
and for Drell-Yan cross sections, so that the convolutions
at finite $Q^2$ depend on the deuteron observable that is
being computed.

For the deuteron wave functions we consider several modern potentials
based on high-precision fits to nucleon--nucleon scattering data.
The models differ primarily in their treatment of the short range
$NN$ interaction, while the long range part of the wave functions is
constrained by the chiral symmetry of QCD and parametrized through
one-pion exchange.
Specifically, the nonrelativistic AV18 \cite{AV18} and CD-Bonn
\cite{CDBonn} $NN$ potential models (which fit around 3,000 data points
in terms of $\approx 40$ parameters), and the more recent relativistic
WJC-1 and WJC-2 \cite{WJC} potentials (which describe almost 4,000
data points in terms 27 and 15 parameters, respectively), provide
wave functions with a representative spread of behaviors in the
low and high momentum regions.  Of these, the CD-Bonn wave function
has the softest momentum distribution, while the WJC-1 wave function
has the hardest, with the others lying between the two.
The differences in the strength of the high-momentum tails of the
wave functions are reflected in differences between the behaviors
of the nuclear corrections at large values of $x$.
Note that the effects of the nuclear smearing corrections are not
suppressed at large $Q^2$, and must be considered at all scales
wherever data at $x \gtrsim 0.3$ are used \cite{CJ10, ACHL09, ARM12}.

\subsubsection{Nucleon off-shell corrections}
\label{sssec:offshell} 

While the effects of the nuclear smearing are relatively well
understood, at least in the sense that they can be directly related to
the properties of the deuteron wave function, the nucleon off-shell
correction in Eqs.~(\ref{eq:qoff}) and (\ref{eq:deltafN}) is much
more uncertain and model dependent.  In the literature a number
of model studies have been performed to estimate the modification
of PDFs in bound nucleons relative to the free nucleon PDFs
\cite{MSToff, KPW94, KMPW95, KP06, GL92, MSTplb, MSS97, Mineo04,
Cloet06}, some of which have been motivated by the original
observation of the nuclear EMC effect \cite{EMC83} (namely, the
deviation of the nuclear to deuterium structure function ratio
from unity).

Some early studies of off-shell corrections to PDFs were based on
spectator quark models \cite{MSToff, KPW94, KMPW95, KP06, MSTplb},
in which the scattering takes place from a quark that is accompanied
by a ``diquark'' system (proton with a quark removed) that is a
spectator to the deep-inelastic collision.  The scattering amplitude
was represented through a quark spectral function characterized by
an ultraviolet momentum cutoff scale $\Lambda$ and an invariant mass
of the spectator system, both of which were fixed by comparing with
the on-shell structure function data.

The effects of nucleon off-shell corrections on global PDF analysis
were explored in the CTEQ6X analysis \cite{CJ10} using a simple
analytic parametrization of the corrections computed in the
relativistic quark spectator model of Ref.~\cite{MSTplb}.
In the subsequent CJ11 analysis \cite{CJ11} a more elaborate off-shell 
model was considered \cite{KP06}, in which the corrections were
related to the change in the nucleon's confinement radius in the
nuclear medium, as well as the average virtuality of the bound
nucleons.  The change in the confinement radius (or nucleon
``swelling'') ranged between 1.5\% and 1.8\%, and the virtuality
of the bound nucleons
  $\langle p^2-M^2 \rangle / M^2 \equiv \int dz\, f^{(\rm off)}(z)$
was independently varied between $-3.6\%$ to $-6.5\%$ for the four
deuteron wave functions discussed above.

Most recently, the CJ12 global analysis \cite{CJ12} further took
into account the correlations between the nucleon swelling and
the deuteron wave function, defining a set of nuclear corrections
that ranged from
mild (for the hardest, WJC-1 wave function \cite{WJC} coupled to a
      small, 0.3\% nucleon swelling) to
strong (for the softest, CD-Bonn wave function \cite{CDBonn} with
      a large, 2.1\% swelling parameter).
The entire range of nuclear corrections was consistent with the
existing experimental data, with each of the CJ12min, CJ12mid
and CJ12max PDF sets giving essentially the same $\chi^2$ values
for the global fit, $\chi^2/{\rm datum} \approx 1.03$.

In the present CJ15 analysis, in order to decrease the model
dependence of the off-shell correction and increase the flexibility
of the fit, we follow the proposal of Kulagin and Petti \cite{KP06}
and employ a phenomenological parametrization with parameters
fitted to data.  From the constraint that the off-shell correction
does not modify the number of valence quarks in the nucleon,
\begin{eqnarray}
\int_0^1 dx\, \delta f^N(x)\, \left[ q(x)-\bar q(x) \right] &=& 0,
\label{eq:norm}
\end{eqnarray}
one can infer that the function $\delta f^N$ must have one
or more zeros in the physical range between $x=0$ and 1.
We can therefore take the off-shell function $\delta f^N$ to be
parametrized by the form
\begin{eqnarray}
\delta f^N
&=& C (x-x_0) (x-x_1) (1+x_0-x),
\label{eq:delffit}
\end{eqnarray}
with the zeros $x_0$ and $x_1$ and normalization $C$ free parameters.
In practice we fit the zero crossing parameter $x_0$ and the
normalization $C$, which then allows the second zero crossing
$x_1$ to be determined from Eq.~(\ref{eq:norm}) analytically.
In Ref.~\cite{KP06} these parameters were constrained by fitting to
ratios of nuclear to deuteron structure function data, for a range
of nuclei up to $^{207}$Pb.  This resulted in a combined nuclear
correction that produced a ratio of deuteron to nucleon structure
functions $F_2^d/F_2^N$ with a shape similar to that for heavy nuclei
\cite{EMC83, Geesaman95}, including an $\approx 1\%$ antishadowing
enhancement in $F_2^d/F_2^N$ at $x \approx 0.1-0.2$.
In contrast, in the present analysis we fit the off-shell parameters
by considering only deuterium cross section data and their interplay
with proton data for a range of processes sensitive to the $d$-quark
PDF.

To test the sensitivity of the fit to the off-shell parametrization,
we also consider as an alternative the model of Ehlers {\it et~al.}
\cite{Ehlers14}, who generalized the quark spectator model employed
in the CJ12 analysis \cite{CJ12} to allow for different off-shell
behaviors of the valence quark, sea quark and gluon distributions.
In previous studies the off-shell corrections were implemented only
for the deuteron $F_2^d$ structure function and in the valence quark
approximation.  The generalized model \cite{Ehlers14}, on the other
hand, which we refer to as the ``off-shell covariant spectator''
(OCS) model, can be applied to observables that are sensitive to
both the valence and sea sectors, such as the deuteron $F_L^d$
structure function or proton--deuteron Drell-Yan cross sections.
More specifically, in the OCS model three masses for the respective
spectator states (``$qq$'' for valence quarks, ``$qq\bar q q$''
for sea quarks, and ``$qqq$'' for gluons) were fitted to the
isoscalar valence, sea quark and gluon PDFs in the free nucleon.
The only free parameter in the model is the rescaling parameter
$\lambda = \partial \log\Lambda^2 / \partial \log p^2$, evaluated
at $p^2=M^2$.  The variable $\lambda$ can then be included as a
parameter in the fit, with errors propagated along with those of
the other leading twist parameters.

Finally, we note that in a purely phenomenological approach adopted
by Martin {\it et al.} \cite{MMSTWW13}, the entire deuterium nuclear
correction is parametrized by a $Q^2$ independent function, without
appealing to physical constraints.  To mock up the effects of Fermi
motion the parametrization includes a logarithm raised to a high
power, $\sim \ln^{20}(x)$, which produces the steep rise in the
$F_2^d/F_2^N$ ratio at high $x$.
In the convolution formula in Eq.~(\ref{eq:genconv}) this effect
arises naturally from the smearing of the nucleon structure function
by the nucleon momentum distribution function $f_{N/d}$.

\section{Data}
\label{sec:data}

The CJ15 PDFs are obtained by fitting to a global database of over
4500 data points from a variety of high energy scattering processes,
listed in Table~\ref{tab:chi2}.
These include deep-inelastic scattering data from
  BCDMS \cite{BCDMS}, SLAC \cite{SLAC}, NMC \cite{NMCp, NMCdop},
  HERA \cite{HERA2}, HERMES \cite{HERMES} and Jefferson Lab
  \cite{Malace, Baillie12, Tkachenko14};
Drell-Yan $pp$ and $pd$ cross sections from fixed target experiments
  at Fermilab \cite{E866};
$W$ \cite{CDF_e, D0_mu, D0_e, CDF_W, D0_W} and $Z$ \cite{CDFZ, D0Z}
  asymmetries, as well as jet \cite{CDFjet2, D0jet2} and $\gamma+$jet
  \cite{D0gamjet} cross sections from the CDF and D\O\ collaborations
  at the Tevatron.
Cuts on the kinematic coverage of the DIS data have been made for
$Q^2 > Q_0^2 = 1.69$~GeV$^2$ and $W^2 > 3$~GeV$^2$, as in the CJ12
analysis \cite{CJ12}. Compared with the CJ12 fit, however, several
new data sets are included in the new analysis.

For DIS, we include the new results from the BONuS experiment
\cite{Baillie12, Tkachenko14} in Jefferson Lab's Hall~B, which
collected around 200 data points on the ratio of neutron to deuteron
$F_2$ structure functions up to $x \approx 0.6$, using a spectator
tagging technique to isolate DIS events from a nearly free neutron
inside a deuterium nucleus \cite{MSS97}.
Unlike all previous extractions of neutron structure from deuterium
targets, which have been subject to large uncertainties in the
nuclear corrections in the deuteron at high $x$ \cite{MT96, KMK09},
the BONuS data provide the first direct determination of $F_2^n$
in the DIS region, essentially free of nuclear uncertainties.

New data sets from the run~II of HERA \cite{Cooper-Sarkar15, HERA2}
and from HERMES on the proton and deuteron structure functions
\cite{HERMES} have become available recently, and are included
in this analysis.  During the fitting process it was noted that
the HERMES data from the highest $Q^2$ bin (bin~``F'' \cite{HERMES})
differed significantly from results from other experiments in the
same kinematic region, and in the final analysis the data in the
$Q^2$ bin~F were not included.  The other DIS data sets are unchanged
from those used in the CJ12 analysis \cite{CJ12}.

For the Drell-Yan data from the E866 experiment \cite{E866} at
Fermilab, following the suggestion in Ref.~\cite{Alekhin06} we
employ a cut on the dimuon cross sections for dimuon masses
$M_{\mu^+\mu^-} > 6$~GeV.  This reduces the number of data
points from 375 to 250 compared to the usual cut of
$M_{\mu^+\mu^-} \gtrsim 4$~GeV, but leads to a significant
reduction in the $\chi^2$/datum for those data.
In previous fits, dimuon data from the E605 Drell-Yan experiment
at Fermilab \cite{E605} were also used.  However, those data
were taken on a copper target and are therefore potentially
subject to nuclear corrections.  Since the nuclear corrections
used in the CJ15 fit pertain only to deuterium targets, we have
chosen not to use the E605 data in this analysis.

Several new data sets from $W$-boson production in $pp$ collisions
at the Tevatron have also recently become available and are included
in the CJ15 fit.  New data from the D\O\ collaboration on muon
\cite{D0_mu} and electron \cite{D0_e} charge asymmetries supersede
previous lepton asymmetry measurements, and remove the tension with
the extracted $W$-boson asymmetries that was evident in our previous
CJ12 analysis \cite{CJ12}.
The new $W$-boson asymmetry data from D\O\ \cite{D0_W} have
about 10 times larger integrated luminosity, and extend over a
larger $W$-boson rapidity range, up to $\approx 3$, than the
earlier CDF measurement \cite{CDF_W}.
While the lepton asymmetry data are more sensitive to PDFs at
small values of $x$, the $W$-boson asymmetry data at large
rapidities generally provide stronger constraints on PDFs
at large $x$ values.

\section{Results}
\label{sec:results}

\begin{figure}[t]
\includegraphics[width=15cm]{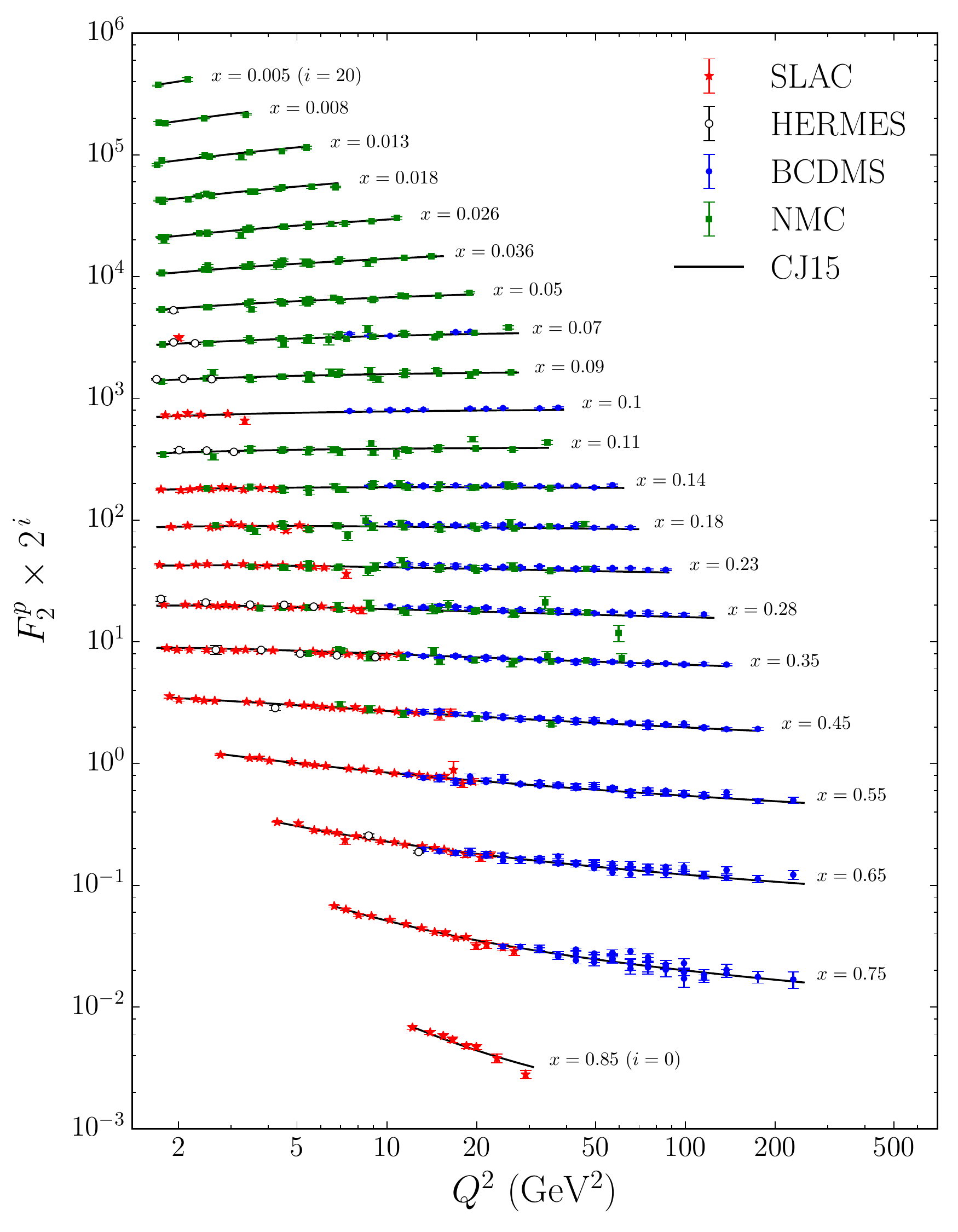}
\caption{Comparison of proton $F_2^p$ structure function data from
	BCDMS \cite{BCDMS},
	SLAC \cite{SLAC},
	NMC \cite{NMCp} and
	HERMES \cite{HERMES} with the CJ15 fit,
	as a function of $Q^2$ for approximately constant $x$.
	The data have been scaled by a factor $2^i$, from $i=0$
	for $x=0.85$ to $i=20$ for $x=0.005$, and the
	PDF uncertainties correspond to a 90\% CL.}
\label{fig:F2p}
\end{figure}

\begin{figure}[t]
\includegraphics[width=15cm]{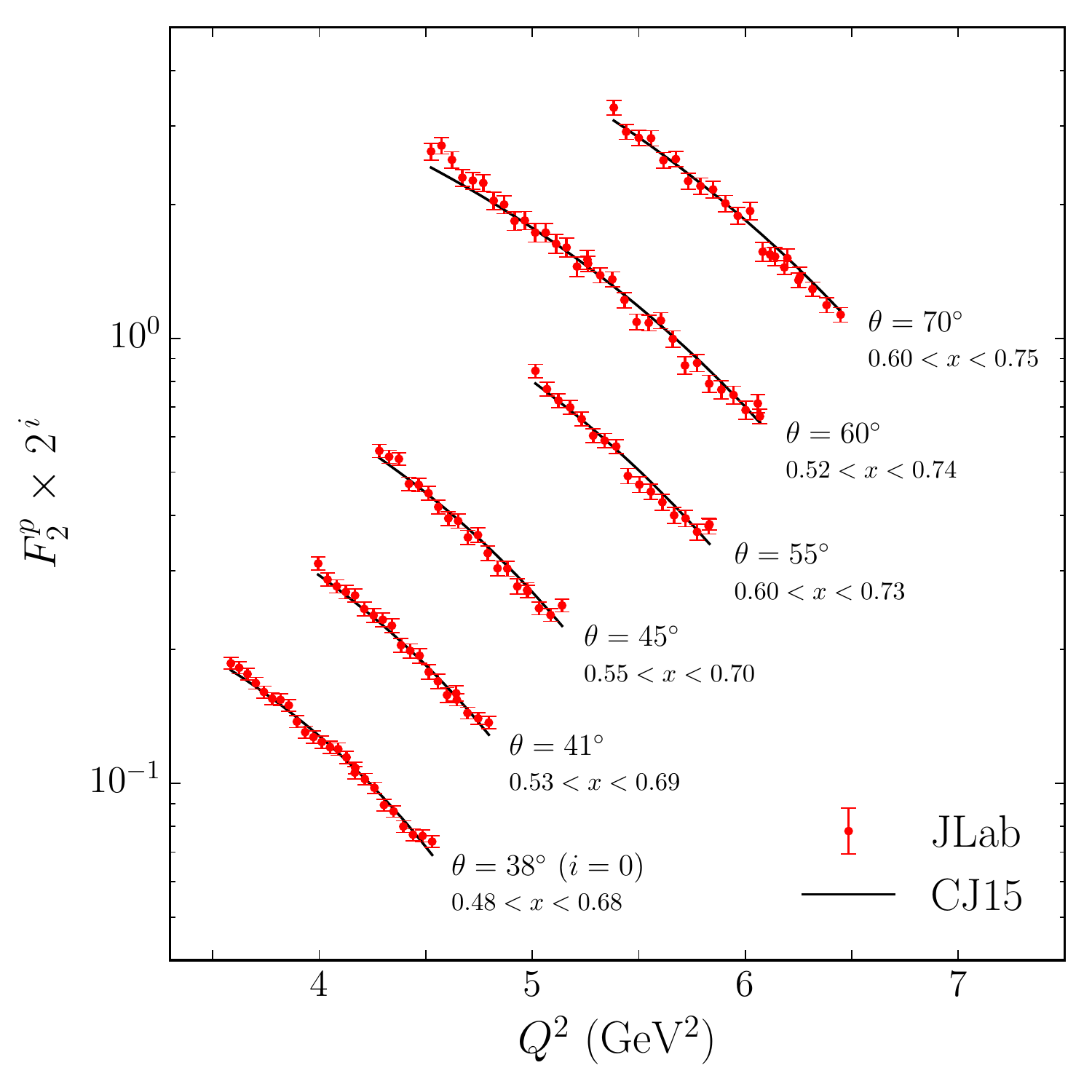}
\caption{Comparison of the proton $F_2^p$ structure function data
	from the E00-116 experiment in Jefferson Lab (JLab) Hall~C
	\cite{Malace} with the CJ15 fit, as a function of $Q^2$ for
	fixed scattering angle $\theta$, with the corresponding $x$
	ranges indicated.
	The data have been scaled by a factor $2^i$, from $i=0$
	for $\theta=38^\circ$ to $i=5$ for $\theta=70^\circ$,
	and the PDF uncertainties correspond to a 90\% CL.}
\label{fig:F2p_JLab}
\end{figure}

\begin{figure}[t]
\includegraphics[width=15cm]{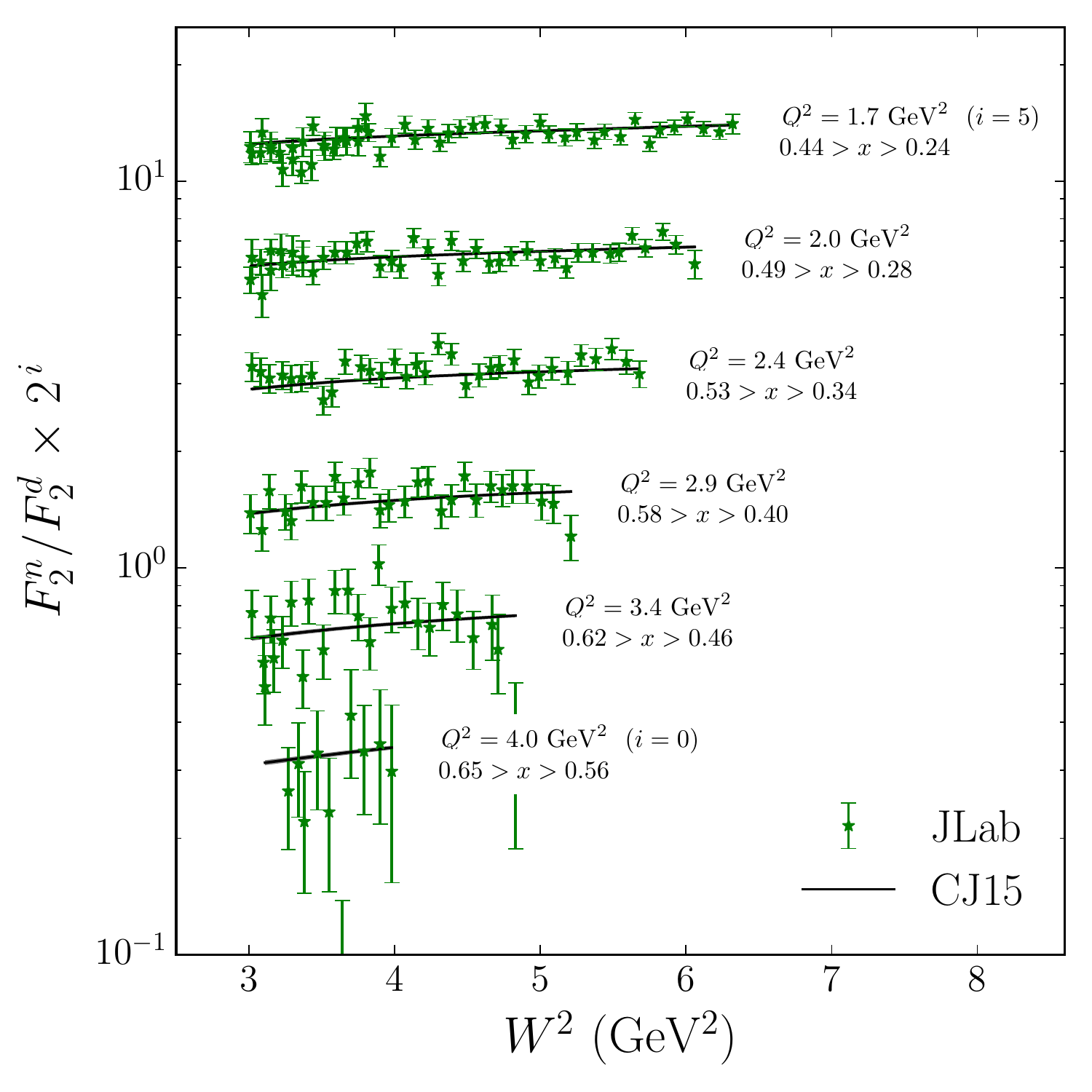}
\caption{Comparison of the $F_2^n/F_2^d$ structure function
	ratio from the BONuS experiment in Jefferson Lab (JLab)
	Hall~B \cite{Tkachenko14} with the CJ15 fit, as a function
	of the invariant mass $W^2$ for fixed $Q^2$, with the
	corresponding $x$ ranges indicated (note $x$ decreases
	with increasing $W^2$).
	The data have been scaled by a factor $2^i$, from $i=0$
	for $Q^2=4.0$~GeV$^2$ to $i=5$ for $Q^2 = 1.7$~GeV$^2$,
	and the PDF uncertainties correspond to a 90\% CL.}
\label{fig:F2n/F2d_BONuS}
\end{figure}

In this section we present the results of our global QCD analysis.
The quality of the fit to the data is illustrated in
Fig.~\ref{fig:F2p}, where the inclusive proton $F_2$ structure
functions from BCDMS \cite{BCDMS}, SLAC \cite{SLAC}, NMC \cite{NMCp}
and HERMES \cite{HERMES} are compared with the CJ15 NLO fit as
a function of $Q^2$ at approximately constant values of $x$.
In Fig.~\ref{fig:F2p_JLab} the Jefferson Lab $F_2^p$ data from the
E00-116 experiment in Hall~C \cite{Malace} are compared with the CJ15
results at fixed scattering angles, with $x$ increasing with $Q^2$.
The more recent data from the BONuS experiment at Jefferson Lab
\cite{Tkachenko14} on the ratio of neutron to deuteron structure
functions, $F_2^n/F_2^d$, are shown in Fig.~\ref{fig:F2n/F2d_BONuS}.
Overall the agreement between the theory and data, over several
decades of $Q^2$ and $x$, is excellent.

The uncertainties on the observables in Figs.~\ref{fig:F2p} --
\ref{fig:F2n/F2d_BONuS} (and on the PDFs throughout this paper,
unless otherwise noted) are computed using Hessian error propagation,
as outlined in Ref.~\cite{CJ12}, with $\Delta\chi^2=2.71$, which
corresponds to a 90\% confidence level (CL) in the ideal Gaussian
statistics.
The corresponding $\chi^2$ values for each of the data sets in
Figs.~\ref{fig:F2p} -- \ref{fig:F2n/F2d_BONuS}, and all other
data used in the fits, are listed in Table~\ref{tab:chi2}.
As well as the main NLO fit, we also include the $\chi^2$ values
for several alternate fits, with different combinations of theory
and data (see below), and an LO fit.
For the central NLO fit, the total $\chi^2$ is $\approx 4700$ for
4542 points, or $\chi^2/{\rm datum} = 1.04$, which is similar to
our previous CJ12 analysis \cite{CJ12}, even though that fit was
to some 500 fewer points.
While the various NLO fits give qualitatively similar $\chi^2$
values, the $\chi^2/{\rm datum}$ for the LO fit ($\sim 1.3$) is
markedly worse.

\subsection{CJ15 PDFs}
\label{ssec:CJ15pdfs}

\begin{figure}[t]
\includegraphics[width=15cm]{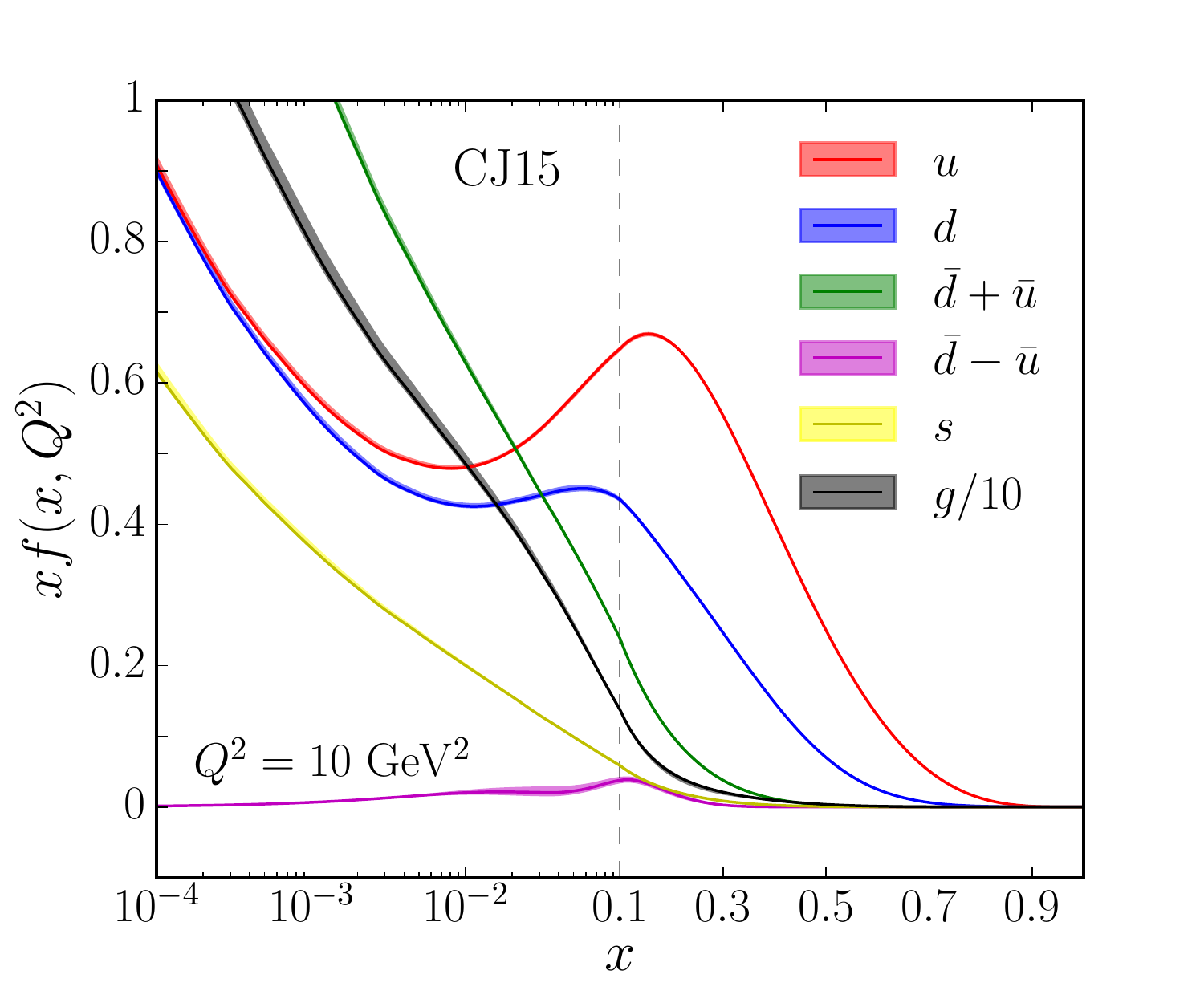}
\caption{Comparison of CJ15 PDFs $xf(x,Q^2)$ for different flavors
	($f=u$, $d$, $\bar d + \bar u$, $\bar d - \bar u$, $s$
	and $g/10$) at a scale $Q^2=10$~GeV$^2$, with 90\% CL
	uncertainty bands.  Note the combined logarithimic/linear
	scale along the $x$-axis.}
\label{fig:pdf}
\end{figure}

The CJ15 PDFs themselves are displayed in Fig.~\ref{fig:pdf} at a
scale of $Q^2=10$~GeV$^2$ for the $u$, $d$, $\bar d + \bar u$,
$\bar d - \bar u$ and $s$ distributions, and the gluon distribution
scaled by a factor 1/10.  The central CJ15 PDFs are determined
using the AV18 deuteron wave function and the nucleon off-shell
parametrization in Eq.~(\ref{eq:delffit}).
The parameter values and their 1$\sigma$ errors for the leading
twist distributions at the input scale $Q_0^2$ are given in
Table~\ref{tab:LTparams}, with the parameters that are listed
without errors fixed by sum rules or other constraints.
(To avoid rounding errors when using these values in numerical
calculations, we give each of the parameter values and their
uncertainties to 5 significant figures.)

The strange quark PDF is assumed in this analysis to be proportional
to the light antiquark sea in the ratio $\kappa = 0.4$ [see
Eq.~(\ref{eq:kappa})].  To test the sensitivity of our fit to the
specific value of $\kappa$, we repeated the analysis varying the
strange to nonstrange quark ratio between 0.3 and 0.5.  Within this
range the total $\chi^2$ spans between 4704 ($\kappa=0.3$) and 4711
($\kappa=0.5$), indicating a very weak dependence on $\kappa$.
This is not surprising given that our analysis does not include any
data sets that are particularly sensitive to the strange-quark PDF.

PDFs for other flavors, such as charm and bottom, are not shown in
Fig.~\ref{fig:pdf}.  The heavy quark distributions are generated
perturbatively through $Q^2$ evolution.
While there has been speculation about nonperturbative or intrinsic
contributions to heavy flavor PDFs, there is currently no evidence
from global analysis of high energy scattering data to suggest that
these are large \cite{PedroIC15}.  Until more conclusive evidence
becomes available, it is reasonable to set these equal to zero.
This is in contrast with the light quark sea, for which a
nonperturbative component at the input scale is essential to
account for the nonzero flavor asymmetry $\bar d-\bar u$.

To study the effect of using the S-ACOT prescription for the $c$
and $b$ quarks, the results for the CJ15 PDFs were compared to
those obtained using the zero-mass variable flavor number (ZMVFS)
scheme.  As expected, the changes to the $u$ and $d$ PDFs were modest,
typically less than 2\%.  On the other hand, an enhancement of up to
40\% was observed for large values of $x \sim 0.4$ for both the
gluon and charm PDF (which are coupled by $Q^2$ evolution).
For the $\bar u$ and $\bar d$ PDFs there was an approximately 5\%
increase near $x \approx 0.1$, followed by a decrease at larger
values of $x$.  However, these effects largely canceled in the
$\bar d/\bar u$ ratio.

The default value for the 5-flavor strong QCD scale parameter
used in our analysis is $\Lambda^{(5)}_{\rm QCD}=0.2268$~GeV,
corresponding to $\alpha_s(M_Z)=0.1180$.  This may be compared
to the world average values quoted by the Particle Data Group,
$\Lambda^{(5)}_{\rm QCD}=(0.2303 \pm 0.0006)$~GeV and
$\alpha_s(M_Z)=0.1185 \pm 0.0006$ \cite{PDG}.
Repeating our standard analysis with $\Lambda^{(5)}_{\rm QCD}$
treated as a free parameter, on the other hand, yields
$\Lambda^{(5)}_{\rm QCD}=(0.230 \pm 0.002)$~GeV and
$\alpha_s(M_Z)=0.1183 \pm 0.0002$,
which are compatible with the PDG results.

\begin{figure}[t]
\includegraphics[width=15cm]{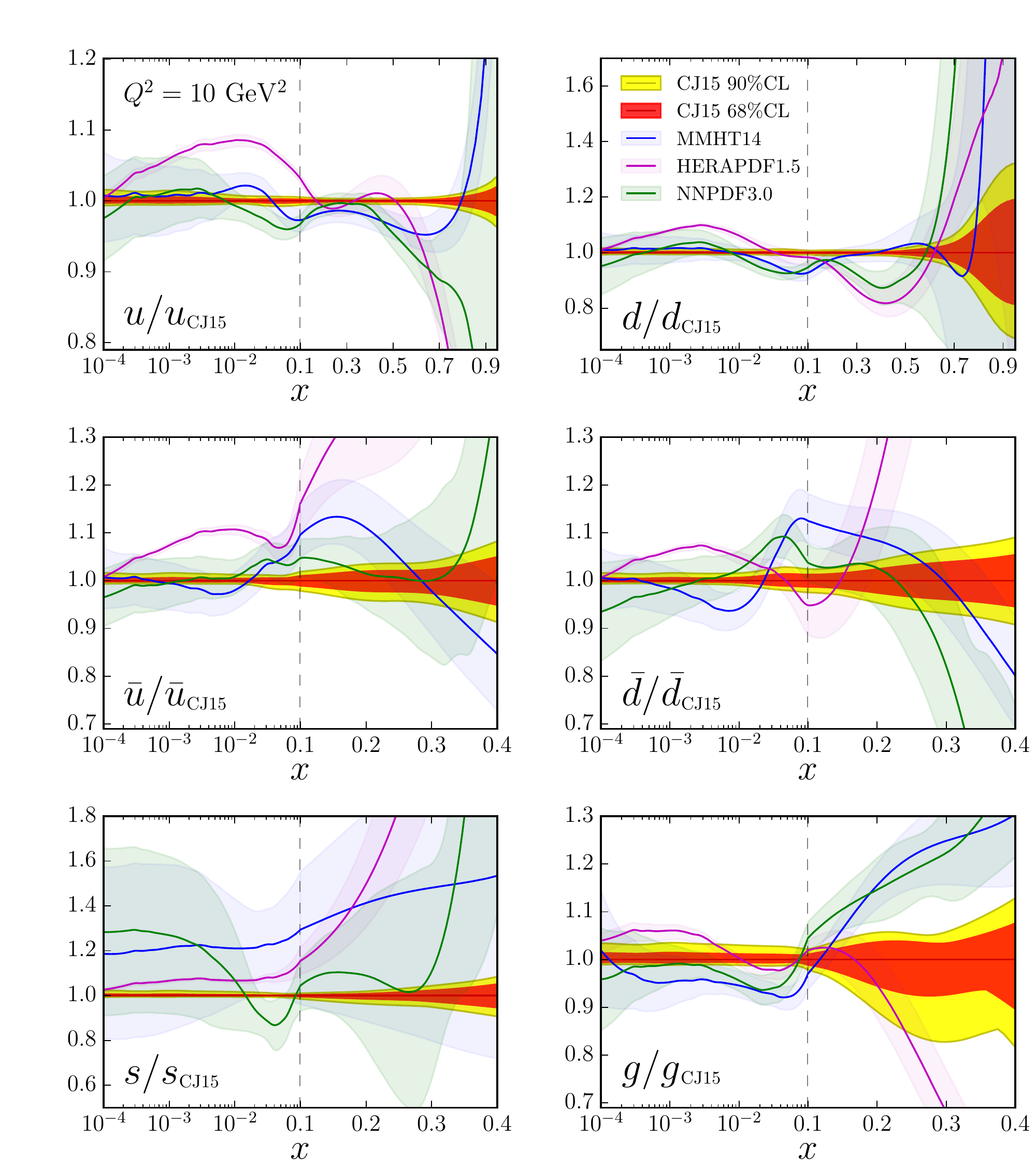}
\caption{Ratio of PDFs to the CJ15 central values for various PDF sets:
	CJ15 with 90\% CL (yellow) and 68\% CL (red)
	uncertainty bands,
	MMHT14 \cite{MMHT14} (blue),
	HERAPDF1.5 \cite{HERAPDF15} (magenta), and
	NNPDF3.0 \cite{NNPDF3.0} (green).
	Note the different scales on the vertical axes used
	for different flavors.}
\label{fig:ratio}
\end{figure}

The CJ15 distributions are compared with PDFs from several recent
representative NLO global parametrizations in Fig.~\ref{fig:ratio},
in the form of ratios to the central CJ15 distributions.  Since
different PDF analyses typically utilize different criteria for
estimating PDF errors, we display the CJ15 errors for the standard
$\Delta\chi^2=1$, or 68\% CL for Gaussian statistics, as well as
with $\Delta\chi^2=2.71$, or 90\% CL.
Generally the MMHT14 \cite{MMHT14} PDF set, which uses a dynamical
tolerance criteria, and the NNPDF3.0 \cite{NNPDF3.0} set have larger
PDF uncertainties than CJ15.  The PDFs uncertainties from HERAPDF1.5
\cite{HERAPDF15} are closer to the CJ15 68\% errors, which may be
expected given that the HERAPDF1.5 analysis only fits HERA data
and uses the $\Delta\chi^2=1$ criterion for generating errors.

For the $u$-quark PDF, the results from different parametrizations
are generally within $5\%$ for $x \lesssim 0.5$, with the
exception of the HERAPDF1.5 set, which is up to $\approx 10\%$
larger at $x \approx 10^{-2}$.  At $x \gtrsim 0.6$, where data
are more limited, there is larger deviation among the PDF sets,
although the uncertainties are correspondingly larger.
A somewhat greater spread between the different parametrizations
is found for the $d$-quark PDF, with the NNPDF3.0 and HERAPDF1.5
results up to 10\%--20\% lower than CJ15 at $x \sim 0.3-0.6$,
while the MMHT14 distribution generally follows CJ15.

As known from previous analyses, the relative uncertainties on
the $d$-quark PDFs are significantly larger than those on the
$u$-quark PDF, especially at large $x$.
For the $\bar u$ and $\bar d$ distributions the results from the
CJ15 fit are similar to those from the MMHT14 and NNPDF3.0 analyses,
while the HERAPDF1.5 fit gives rather different results beyond
$x \approx 0.1-0.2$.  Note that the $\bar d/\bar u$ ratio is most
strongly constrained by the E866 Drell-Yan $pp$ and $pd$ scattering
data.

For the strange quark PDF, the uncertainties in CJ15 are somewhat
smaller than for MMHT14 and NNPDF3.0.  This is mostly due to the
fact that the CJ15 $s$-quark PDF is assumed to scale with the
light antiquark sea in the ratio $\kappa=0.4$, while other analyses
attempt to constrain $s$-quark PDF from neutrino data, which
typically have much larger uncertainties.
The errors on the gluon distribution in the MMHT14 and NNPDF3.0 fits
are comparable to the 90\% CL CJ15 errors, while the HERAPDF1.5
uncertainties are similar to the 68\% CL CJ15 results.
Uncertainties in other modern PDF analyses, such as CT14 \cite{CT14},
JR14 \cite{JR14} or ABM11 \cite{ABM11}, are generally between the
representative sets illustrated in Fig.~\ref{fig:ratio}.

\subsection{Impact of new data sets and interplay of proton and
	nuclear data}
\label{sec:impact-interplay}

The impact of the combined HERA run~I and II inclusive proton DIS
cross sections \cite{HERA2} has been discussed recently in
Ref.~\cite{HMMT16}, with particular focus on the small-$x$ region.
Compared to only using data from run~I, we also find rather stable
PDF central values.  In the large-$x$ region, the improvement in
the PDF uncertainty is
  $\sim 10\%$ for the $u$ distribution at $x \approx 0.05-0.7$,
  $\sim 5\%$ for the $d$ distribution at $x \approx 0.05-0.4$
(and slightly less for the $d/u$ ratio because of anticorrelations
between these), and
 $\sim 5\%$ for the gluon PDF at $x \approx 0.05-0.5$.
The influence of the HERMES data on the proton and deuteron $F_2$
structure functions is less pronounced.  These data induce a minor
reduction, of less than 5\%, in the uncertainty on the $u$ and $d$
PDFs at $x \lesssim 0.2$, which shrinks to less than 2\% in the
$d/u$ ratio.  This is due in part to the limited number of data
points surviving our cuts, and the relatively large systematic
errors compared with the other DIS data sets.

The most notable impact of the new data sets on the CJ15 fit is from
the high-precision D\O\ data on the reconstructed $W$ charge asymmetry
\cite{D0_W}.  These data allow us to simultaneously reduce the
uncertainty on the $d$-quark PDF at $x \gtrsim 0.4$ by $\sim 50\%$
and fit the off-shell correction $\delta f^N$ in Eq.~(\ref{eq:qoff}).
This is possible only in the context of a global fit, by considering
simultaneously the $W$ asymmetry and deuteron DIS structure functions.
If the $d$-quark PDF in the free nucleon can be determined with
sufficient precision, the deuteron DIS data can then be used to
constrain the nuclear corrections, and in particular, for a given
deuteron wave function, the off-shell correction $\delta f^N$.
In principle, the Jefferson Lab BONuS data \cite{Baillie12, Tkachenko14}
on quasi-free neutrons can play an analogous role.  Unfortunately,
the statistics and kinematic reach at large $x$ of the current data
make this difficult, although future data from several planned
experiments \cite{MARATHON, BONUS12, SOLID} at the energy-upgraded
Jefferson Lab are expected to cover the required range in $x$ with
high precision.

This interplay between the proton and nuclear observables is
already evident at the $\chi^2$ level from Table~\ref{tab:chi2}.
When fitting data without including any nuclear corrections,
significantly worse $\chi^2$ values are obtained for the SLAC
deuteron $F_2$ measurement and the D\O\ $W$ asymmetry in particular,
increasing by 131 units over 582 points and 68 units over 14 points,
respectively.  
Similar results are obtained when using the OCS model for the
off-shell corrections instead of the parametrization in
Eq.~(\ref{eq:qoff}).
Without nuclear corrections, a strong tension exists between the
$d$-quark PDF constrained by one or the other of these observables.
This is the first direct indication from a global PDF fit of the
necessity of nuclear corrections, and opens the way for utilizing
proton data to study the dynamics of partons in nuclei
\cite{Accardi13,Accardi16}.

After including nuclear corrections, the D\O\ $W$ asymmetry data
can be fitted with $\chi^2/{\rm datum} \approx 1$, and the SLAC
deuteron $F_2$ data gives an even smaller $\chi^2$ than that
obtained when fitting with no corrections and no D\O\ data.
The tension between these data sets is therefore completely
removed by accounting for nuclear effects.
Interestingly, the fit without nuclear corrections improves the
$\chi^2$ for the D\O\ muon asymmetry data \cite{D0_mu}, but gives
a worse fit to the D\O\ electron asymmetry data \cite{D0_e}.
Although less dramatically, nuclear corrections also improve
the fit to the E886 $pd$ Drell-Yan data.

Overall, it is encouraging that such a relatively simple
parametrization for the nucleon off-shell corrections as used
in this analysis is able to capture most of the effects in DIS
and Drell-Yan observables, in which both valence and sea quarks
play a role.
With the upcoming data from the SeaQuest experiment at Fermilab
\cite{SeaQuest} (and in the future from JPARC, as well as from
dilepton, $W$ and $Z$ boson measurements in $pd$ collisions at RHIC),
separation of off-shell effects in the valence and sea quark
sectors may become feasible.

\subsection{Nuclear corrections at large $x$}
\label{ssec:largex}

As observed in Fig.~\ref{fig:ratio}, the uncertainty on the $d$-quark
distribution at large $x$ values ($x \gtrsim 0.3$) is generally much
larger compared with that on the $u$-quark PDF.
This reflects the considerably greater quantity of high-precision
proton $F_2$ structure function data, which, because of the larger
charge on the $u$ quark, is at least an order of magnitude more
sensitive to the $u$-quark PDF than to the $d$.
Traditionally, stronger constraints on the $d$-quark PDF have been
sought from inclusive DIS from the neutron, in which the roles of
the $u$ and $d$ quark are reversed relative to the proton.  However,
the absence of free neutron targets has meant that neutron structure
information has had to be extracted from measurements on deuterium
nuclei.  Unfortunately, at high values of $x$ ($x \gtrsim 0.5$) bound
state effects in the deuteron become important, and uncertainties in
their computation become progressively large with increasing $x$.

\begin{figure}[t]
\includegraphics[width=15cm]{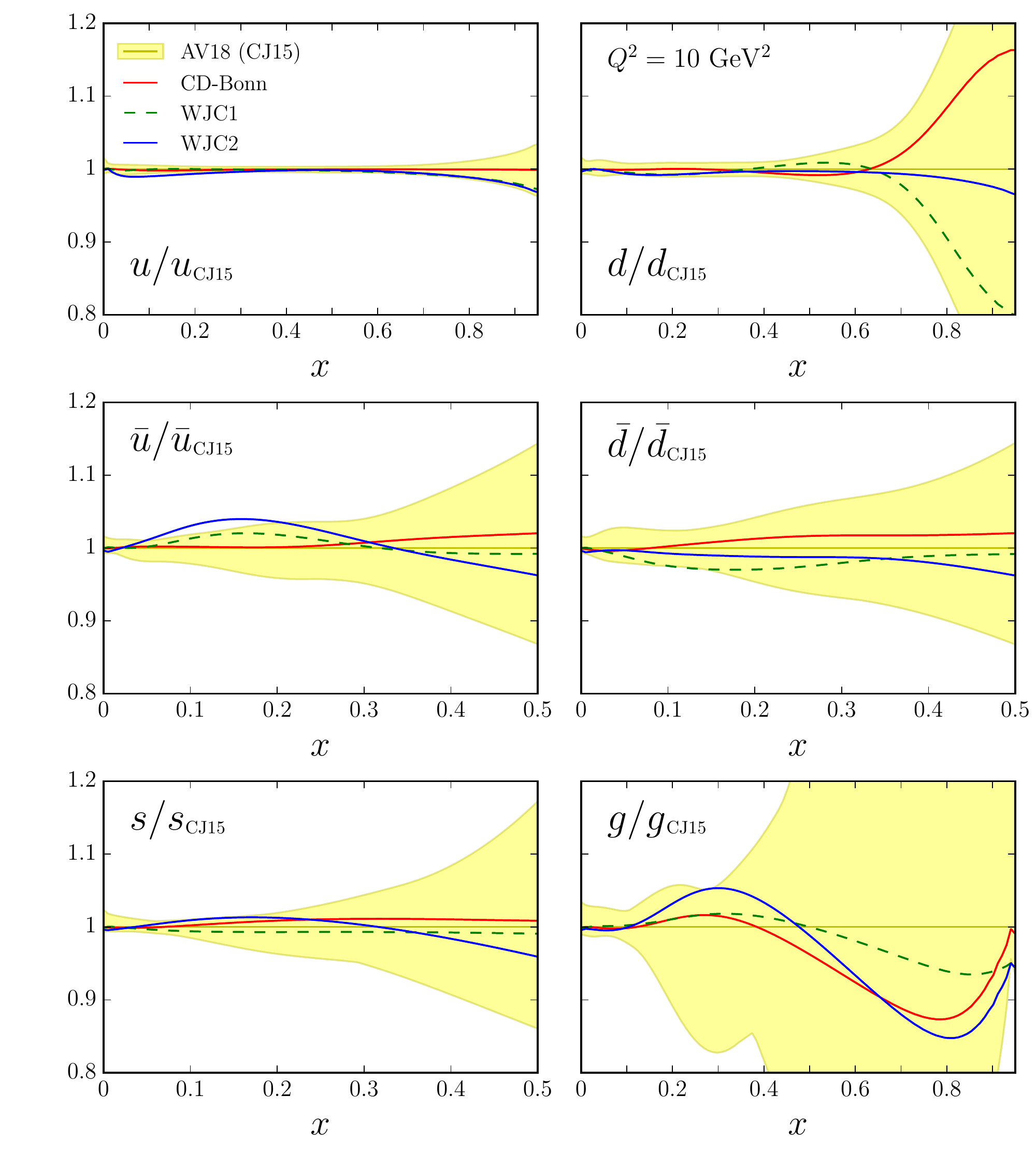}
\caption{Ratio of PDFs fitted using various deuteron wave function models
	to the CJ15 PDFs (which use the AV18 deuteron wave function):
	CD-Bonn (solid red curves),
	WJC-1 (dashed green curves),
	WJC-2 (solid blue curves).
	The CJ15 PDFs (yellow band) correspond to a 90\% CL,
	and the off-shell parametrization (\ref{eq:delffit})
	is used for all cases.}
\label{fig:ratio_wfn}
\end{figure}

The effects of nuclear corrections on the PDFs are illustrated in
Fig.~\ref{fig:ratio_wfn}, where fits using several different deuteron
wave function models are compared.  The distributions are displayed
relative to the central CJ15 PDFs which use the AV18 deuteron wave
function.  All the fits employ the phenomenological nucleon off-shell
parametrization in Eq.~(\ref{eq:delffit}), with the parameters given
in Table~\ref{tab:other_params} for the AV18 deuteron wave function.
The results using the CD-Bonn wave function are very similar to
those for the AV18 wave function, while the WJC-1 and WJC-2 models
lead to slightly larger differences in some of the PDFs shown in
Fig.~\ref{fig:ratio_wfn}.  On the other hand, the $\chi^2$ values
for the AV18, CD-Bonn and WJC-2 models are almost indistinguishable,
with the WJC-1 model giving a marginally larger total $\chi^2$
(4714 instead of 4700).
This suggests that, for the most part, the nucleon off-shell
parametrization in Eq.~(\ref{eq:delffit}) is sufficiently flexible
to compensate for changes induced by these wave functions.
For the WJC-1 model it is a little more difficult for the off-shell
corrections to compensate for this wave function's harder momentum
distribution (relative to the other models) within the constraints
of Eq.~(\ref{eq:delffit}), and this leads to a slightly worse 
overall fit. Observables separately sensitive to the nucleon offshellness, 
such as deuteron target DIS with a large momentum detected spectator 
would be needed to separate these two effects.

As expected, the variations due to the nuclear models have the
largest effects in the $d$-quark distribution, which is less
constrained by proton data and hence more sensitive to uncertainties
in the extracted neutron structure function.  The spread in the
$d$-quark PDF at $x=0.8$ is $\approx 20\%$ between the four wave
functions.  The variations for the AV18 and CD-Bonn wave functions
are generally within the $\Delta\chi^2=1$ CL, while for the WJC-2
model the $u$ and $\bar u$ distributions show the biggest deviations,
in the vicinity of $x \sim 0.1-0.2$.
For the WJC-1 deuteron model, the $d$-quark PDF is suppressed at
high $x$ relative to that in the other models, which correlates
with the harder smearing function $f_{N/d}$ at large values of the
nucleon light-cone momentum $y$ and hence a larger $F_2^d/F_2^N$
ratio at high $x$.
As already noted in the CJ11 analysis \cite{CJ11}, there is an
anti-correlation between the behavior of the $d$-quark distribution
at large $x$ and the gluon PDF.  In fact, using the WJC-1 wave
function leads to a slight decrease in all the quark PDFs at
high $x$ (within the range constrained by the data), while the
gluon PDF has the opposite trend.  The spread in the gluon PDF
is $\lesssim 10\%$ for $x < 0.7$, although beyond $x \approx 0.3$
the gluon distribution has a very large uncertainty.

\begin{figure}[t]
\includegraphics[width=15cm]{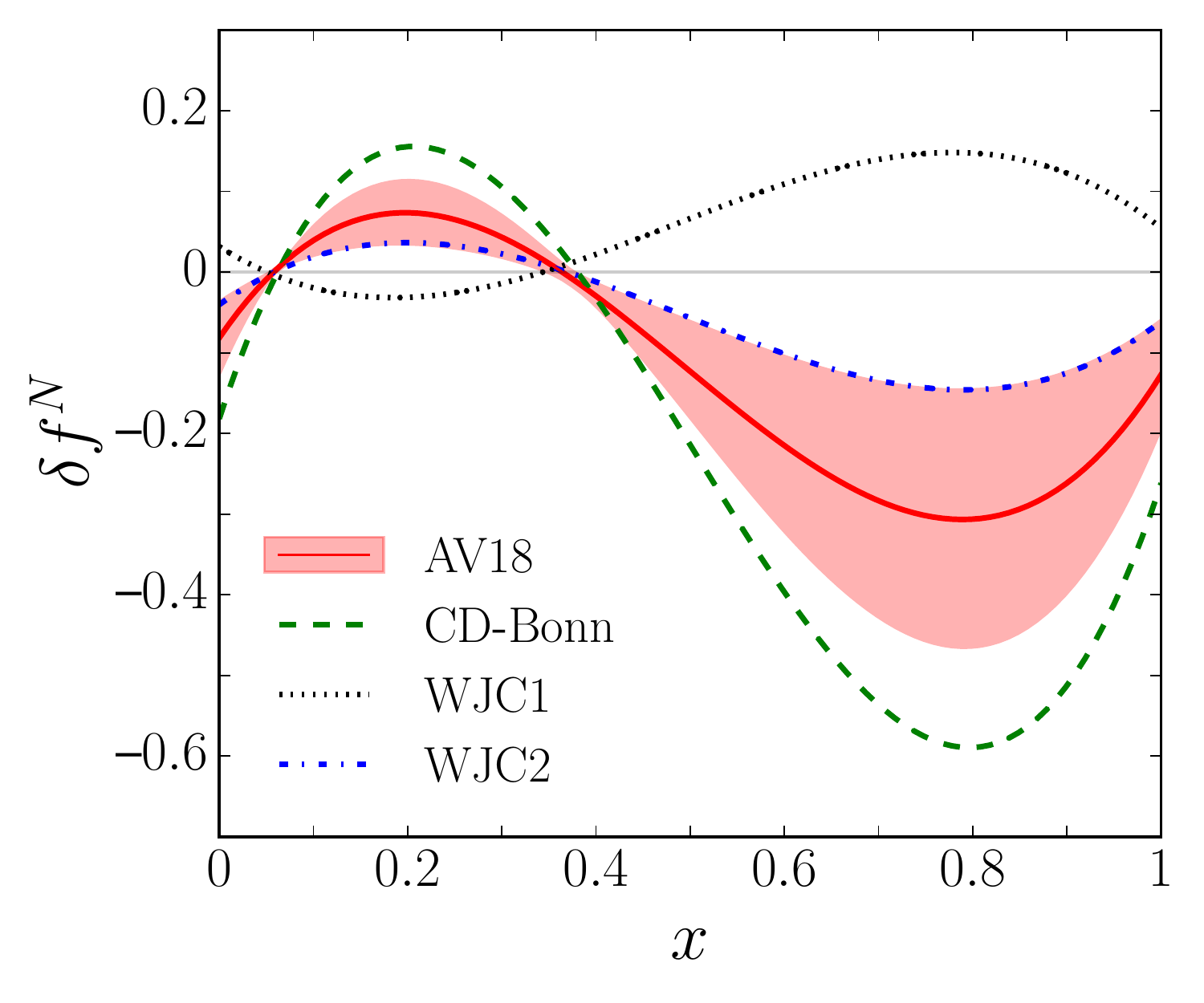}
\caption{Fitted nucleon off-shell correction $\delta f^N$ for
	the parametrization in Eq.~(\ref{eq:delffit}), using the
	AV18 (solid red curve with 90\% CL uncertainty band),
	CD-Bonn (dashed green curve),
	WJC-1 (dotted black curve) and
	WJC-2 (dot-dashed blue curve) wave functions.}
\label{fig:off_shell}
\end{figure}

Note that while in Fig.~\ref{fig:ratio_wfn} the same functional
form from Eq.~(\ref{eq:delffit}) is used for all fits, the off-shell
parameters are refitted for each different deuteron wave function
model, thereby absorbing most of the effect of the varying strength
of the nucleon's momentum distribution tail.
The fitted off-shell functions $\delta f^N$ are shown in
Fig.~\ref{fig:off_shell} for the four wave function models
considered.  The off-shell corrections for the AV18, CD-Bonn
and WJC-2 models have similar shapes: quite small at low $x$,
but more negative at larger $x$, with magnitude peaking at
$x \sim 0.8$.
The function $\delta f^N$ for these models has zero crossings
at $x = x_1 \approx 0.05$ and $x = x_0 \approx 0.35$.
For the WJC-1 model, on the other hand, the off-shell function
is somewhat orthogonal to the others, becoming negative at
lower $x$ values, and positive at higher $x \gtrsim 0.4$.

\begin{figure}[t]
\includegraphics[width=15cm]{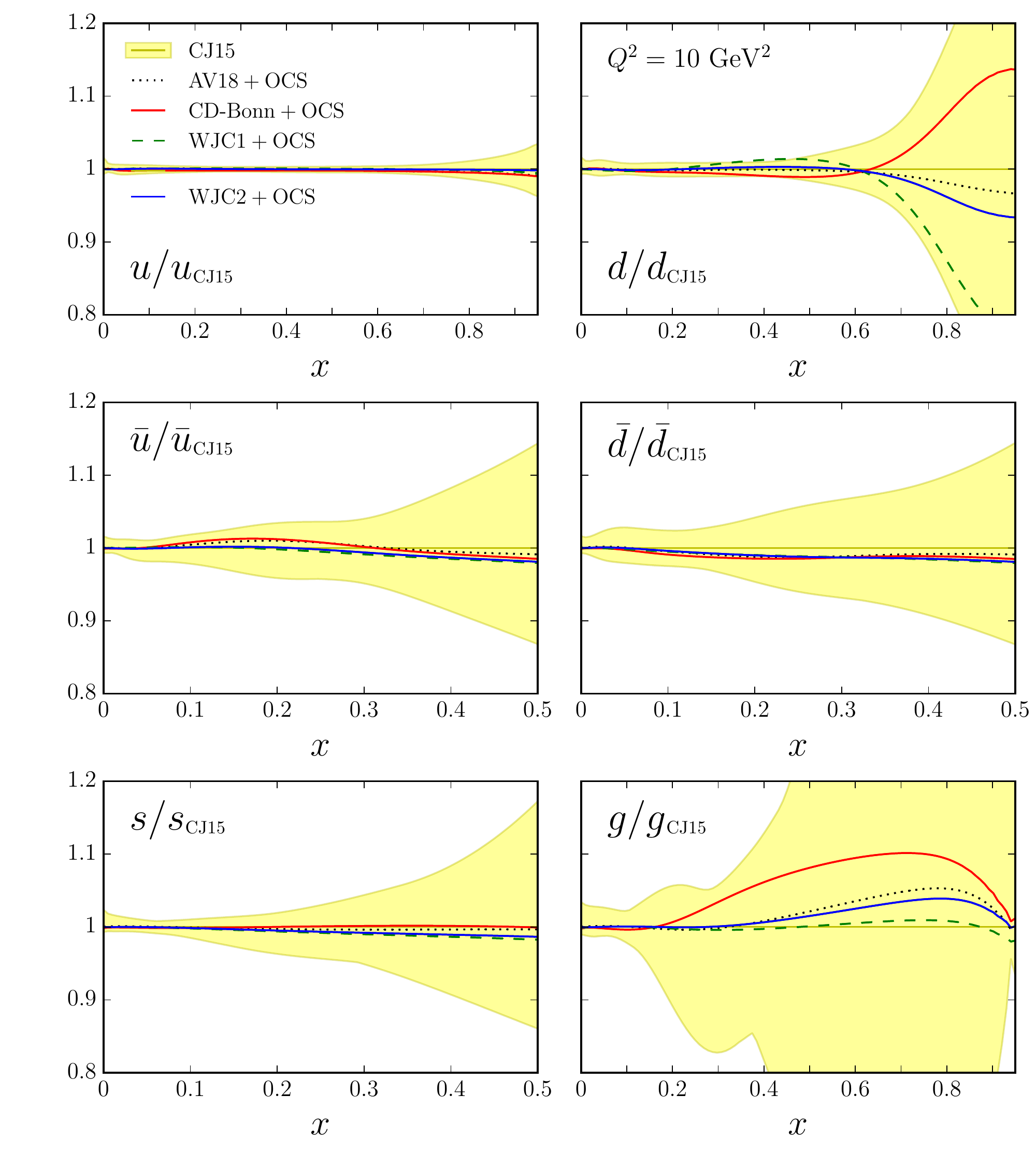}
\caption{Ratio of PDFs computed using the off-shell covariant spectator
	(OCS) model and different deuteron wave functions to the CJ15
	PDFs (which use the off-shell parametrization (\ref{eq:delffit})
	and the AV18 deuteron wave function):
	OCS model with the AV18 wave function (black dotted curves),
	CD-Bonn (solid red curves),
	WJC-1 (dashed green curves), and
	WJC-2 (solid blue curves).
	The yellow band shows the 90\% CL for the CJ15 fit.}
\label{fig:ratio_off}
\end{figure}

To test the sensitivity of the fit to the choice of off-shell
model, we also consider the more microscopic OCS model for
$\delta f^N$ discussed in Sec.~\ref{sssec:offshell}.
The rescaling parameter $\partial \log\Lambda^2 / \partial \log p^2$
evaluated at $p^2=M^2$ is then included as a parameter in the fit,
with errors propagated along with those of the other fit parameters
in Tables~\ref{tab:LTparams} and \ref{tab:other_params}.
The results of the fit using the OCS model are displayed in
Fig.~\ref{fig:ratio_off} for various PDFs as ratios to the
central CJ15 PDFs (computed using the off-shell parametrization
(\ref{eq:delffit}) and the AV18 deuteron wave function).
For most of the PDFs the effects of using the more restrictive
OCS model are relatively small and generally within the
$\Delta\chi^2=1$ bands for all wave function models.
The largest effects are for the $d$-quark distribution,
where the results with the WJC-1 wave function show greater
deviation at intermediate and large $x$ values, suggesting
that the hard tail of its momentum distribution may be more
difficult to accommodate also within the OCS model.
The overall $\chi^2$ values for all wave functions are similar
to those of the main CJ15 fit, with differences in $\chi^2$/datum
appearing only in the third decimal place.

We should note, however, that the off-shell correction term
$\delta f^N$, or even the off-shell PDF $\widetilde{q}^N$ in
Eq.~(\ref{eq:qoff}), alone is nonphysical.  Only the convolution
of $\widetilde{q}^N$ with the smearing function $f_{N/d}$ in
Eq.~(\ref{eq:genconv}) corresponds to the physical deuteron
parton distribution (or structure function), and the two
corrections (deuteron wave function and nucleon off-shell)
must always be considered together.
Since the off-shell correction is fitted, changes in deuteron
wave function can in practice be compensated by a corresponding
change in the off-shell parameters, to the extent allowed by the
specific choice of wave function and off-shell parametrization.

This is clearly illustrated in Fig.~\ref{fig:F2dN}, where the
deuteron to nucleon $F_2^d/F_2^N$ ratio is shown for the four
different wave functions considered, and the 3-parameter
off-shell parametrization in Eq.~(\ref{eq:delffit}).
Remarkably, the structure function ratio is almost identical
for the AV18, CD-Bonn and WJC-2 models.  The slightly larger
differences with the WJC-1 result reflecting the observations
in Figs.~\ref{fig:ratio_wfn} -- \ref{fig:ratio_off} above,
but even in this case the ratio is within the $\Delta\chi^2=1$
uncertainty band.

\begin{figure}[t]
\includegraphics[width=15cm]{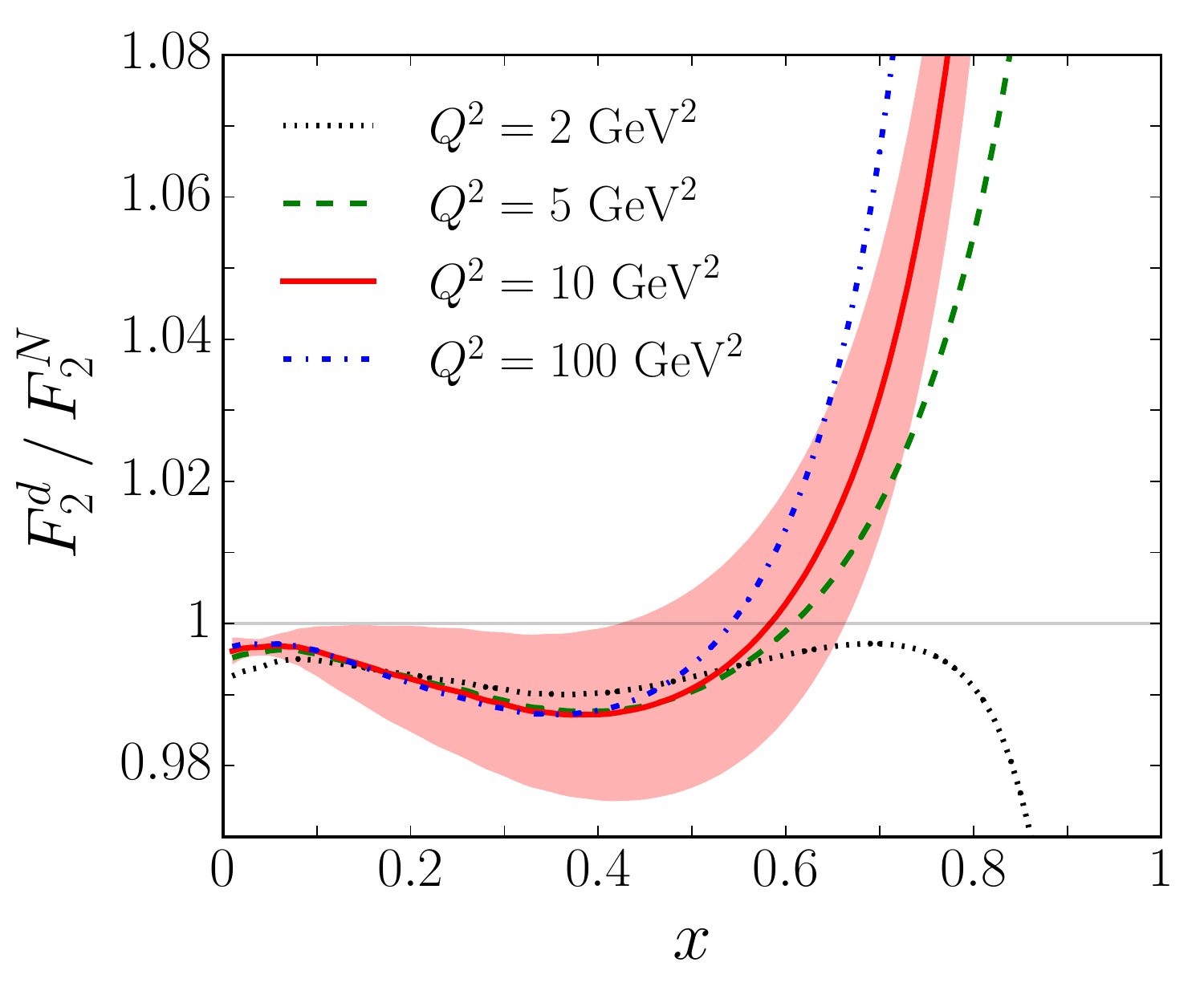}
\caption{Ratio of deuteron to isoscalar nucleon structure functions
	$F_2^d/F_2^N$ computed from the CJ15 PDFs for different
	values of $Q^2$:
	2~GeV$^2$ (dotted black curve),
	5~GeV$^2$ (dashed green curve),
	10~GeV$^2$ (solid red curve with 90\% CL uncertainty band) and
	100~GeV$^2$ (dot-dashed blue curve).}
\label{fig:F2dN_Q2}
\end{figure}

In addition to the important role played by nuclear corrections
in $F_2^d$ at large values of $x$, the effects of finite-$Q^2$
corrections are also significant, especially at low $Q^2$.
In an earlier study \cite{CJ10}, a nontrivial interplay was
observed between the kinematic TMCs and the dynamical higher
twist corrections parametrized in Eq.~(\ref{eq:F2ht}).
The impact of the finite-$Q^2$ corrections on the $F_2^d/F_2^N$
ratio is illustrated in Fig.~\ref{fig:F2dN_Q2} for the CJ15 fit,
for $Q^2$ between 2~GeV$^2$ and 100~GeV$^2$.
The rise in the ratio at large $x$ is fastest at the highest
$Q^2$ value, and becomes less steep with decreasing $Q^2$.
The general shape remains independent of the scale for
$Q^2 \gtrsim 5$~GeV$^2$; however, a dramatic change occurs at
$Q^2 \sim 2$~GeV$^2$, where $F_2^d/F_2^N$ rises slowly until
$x \approx 0.75$, before abruptly turning down for larger $x$.
This behavior arises from the interplay between the target mass
and higher twist corrections to the free and bound nucleon
structure functions, and the $Q^2$ dependent corrections to
the smearing functions $f_{N/d}$ at finite values of $Q^2$.

\begin{figure}[t]
\includegraphics[width=15cm]{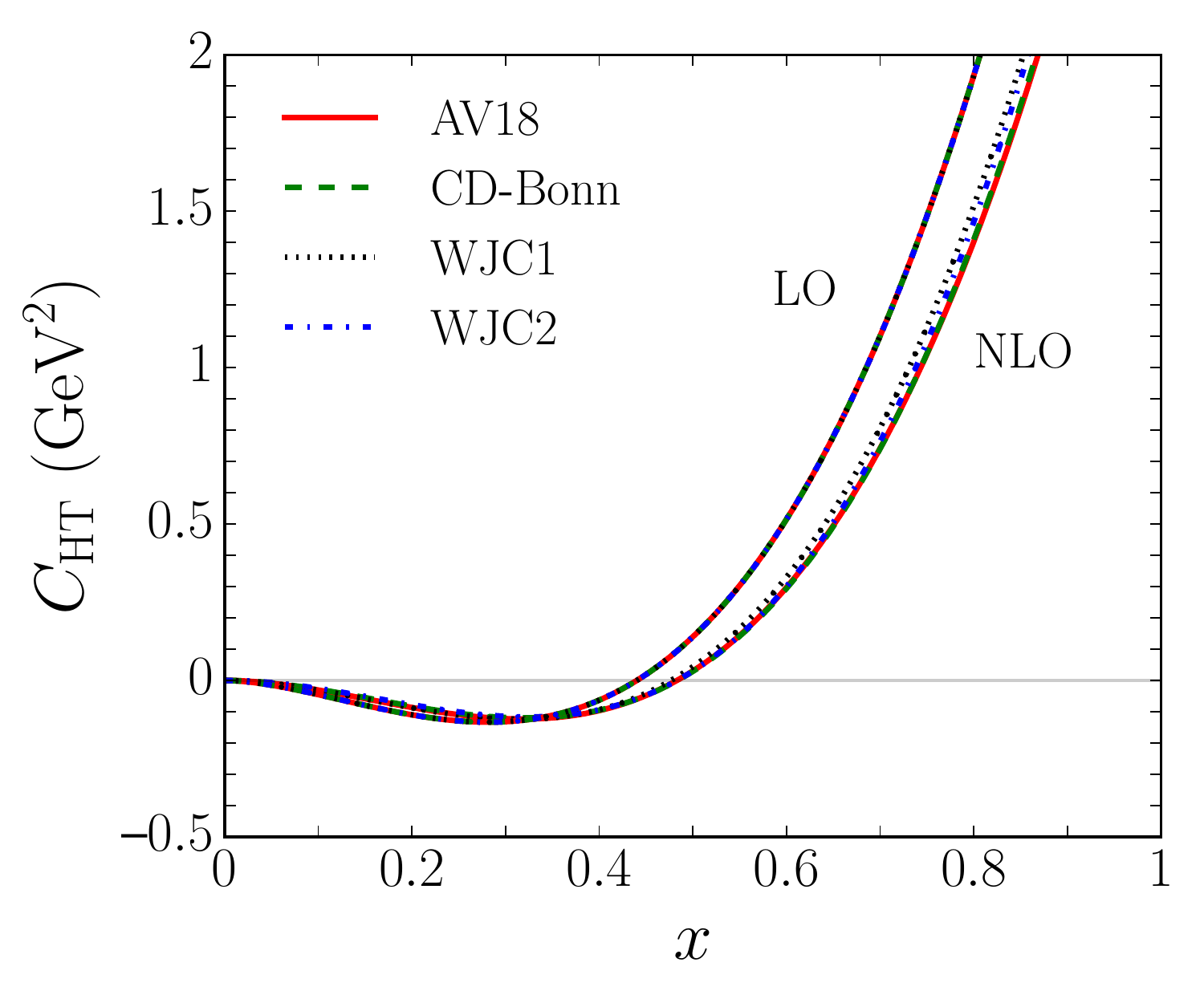}
\caption{Fitted higher twist function $C_{\rm HT}$ from
	Eq.~(\ref{eq:C_ht}), in units of GeV$^2$, for different
	deuteron wave function models.  The higher twist term for
	the CJ15 NLO fit is compared with the corresponding term
	in the LO fit.  The 90\% CL uncertainty band is barely
	visible and is not shown here.}
\label{fig:Cht}
\end{figure}

For the standard TMC prescription adopted in this analysis,
based on the operator product expansion \cite{GP76},
the fitted higher twist coefficient function $C_{\rm HT}$
in Eq.~(\ref{eq:C_ht}) is plotted in Fig.~\ref{fig:Cht},
with the parameters given in Table~\ref{tab:other_params}
for the CJ15 fit.
The coefficient displays the characteristic rise at large
values of $x$ observed in previous higher twist extractions,
and is almost completely independent of the deuteron wave
function model over the entire range of $x$ considered.
For LO fits, the higher twist function also absorbs part of
the missing NLO contributions, resulting in higher values of
the $C_{HT}$ coefficient at large $x$, as was observed also
in Refs.~\cite{BB08, MRST03}.

\subsection{$d/u$ ratio}
\label{ssec:du}

\begin{figure}[t]
\includegraphics[width=15cm]{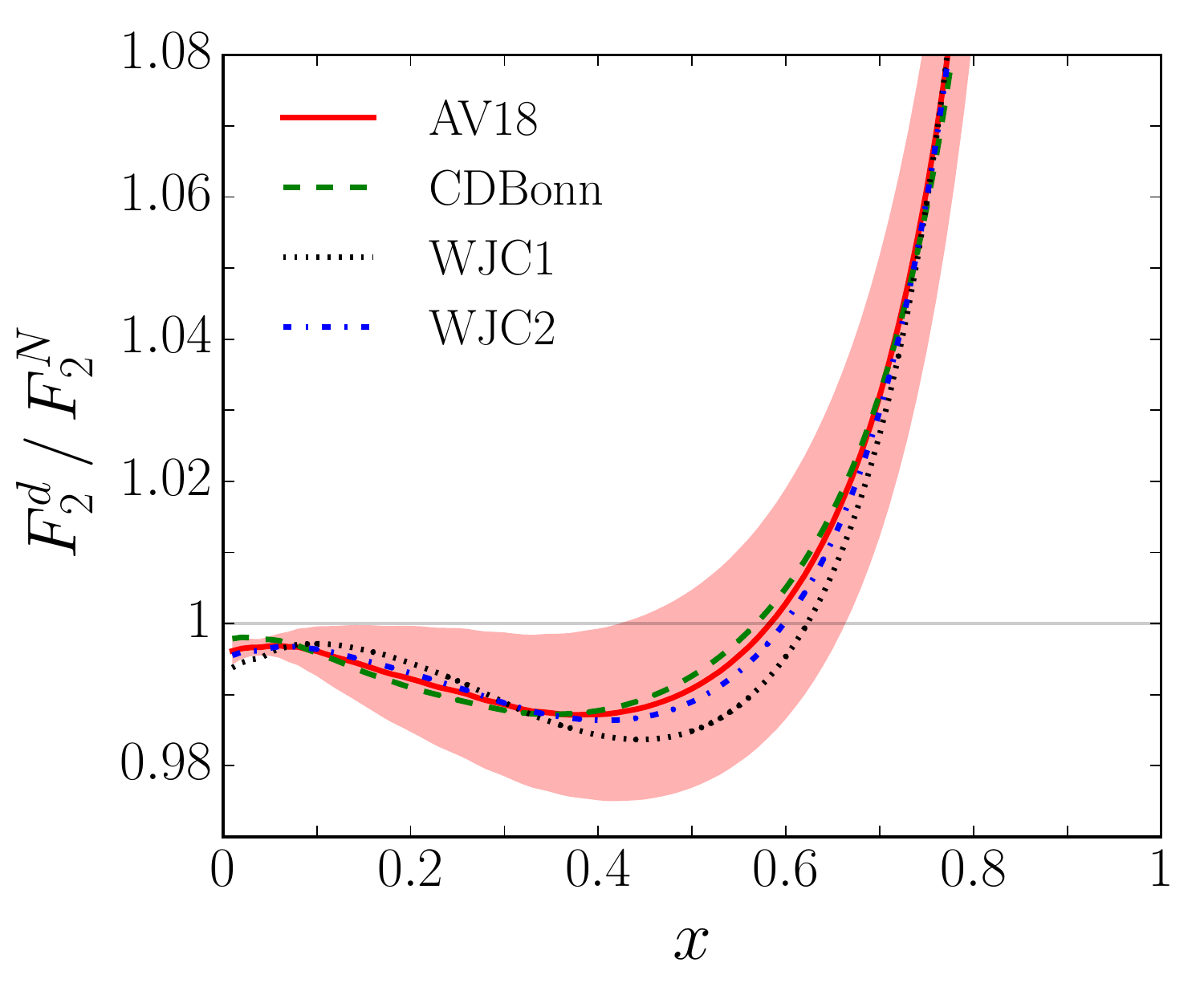}
\caption{Ratio of deuteron to isoscalar nucleon structure functions
	$F_2^d/F_2^N$ for different deuteron wave function models
	at $Q^2=10$~GeV$^2$:
        AV18 (solid red curve with 90\% CL uncertainty band),
        CD-Bonn (dashed green curve),
        WJC-1 (dotted black curve) and
        WJC-2 (dot-dashed blue curve).}
\label{fig:F2dN}
\end{figure}

The nuclear and finite-$Q^2$ corrections that manifest themselves
in the $F_2^d/F_2^N$ ratio as observed in Figs.~\ref{fig:F2dN} and
\ref{fig:F2dN_Q2} directly translate into modifying the behavior
of the $d/u$ PDF ratio at large $x$.  Our previous analyses
\cite{CJ10, CJ11, CJ12} have made detailed studies of the
relationship between the size of the nuclear corrections in
the deuteron and the shape and $x \to 1$ limit of $d/u$.
For the CJ12 PDFs \cite{CJ12}, three sets of nuclear corrections
were considered, corresponding to mild, medium and strong nuclear
corrections, and referred to as ``CJ12min'', ``CJ12mid'' and
``CJ12max'', respectively.  Each of these sets was consistent
with the available data constraints, and provided a convenient
way to explore the nuclear effects on various observables.

\begin{figure}[t]
\includegraphics[width=15cm]{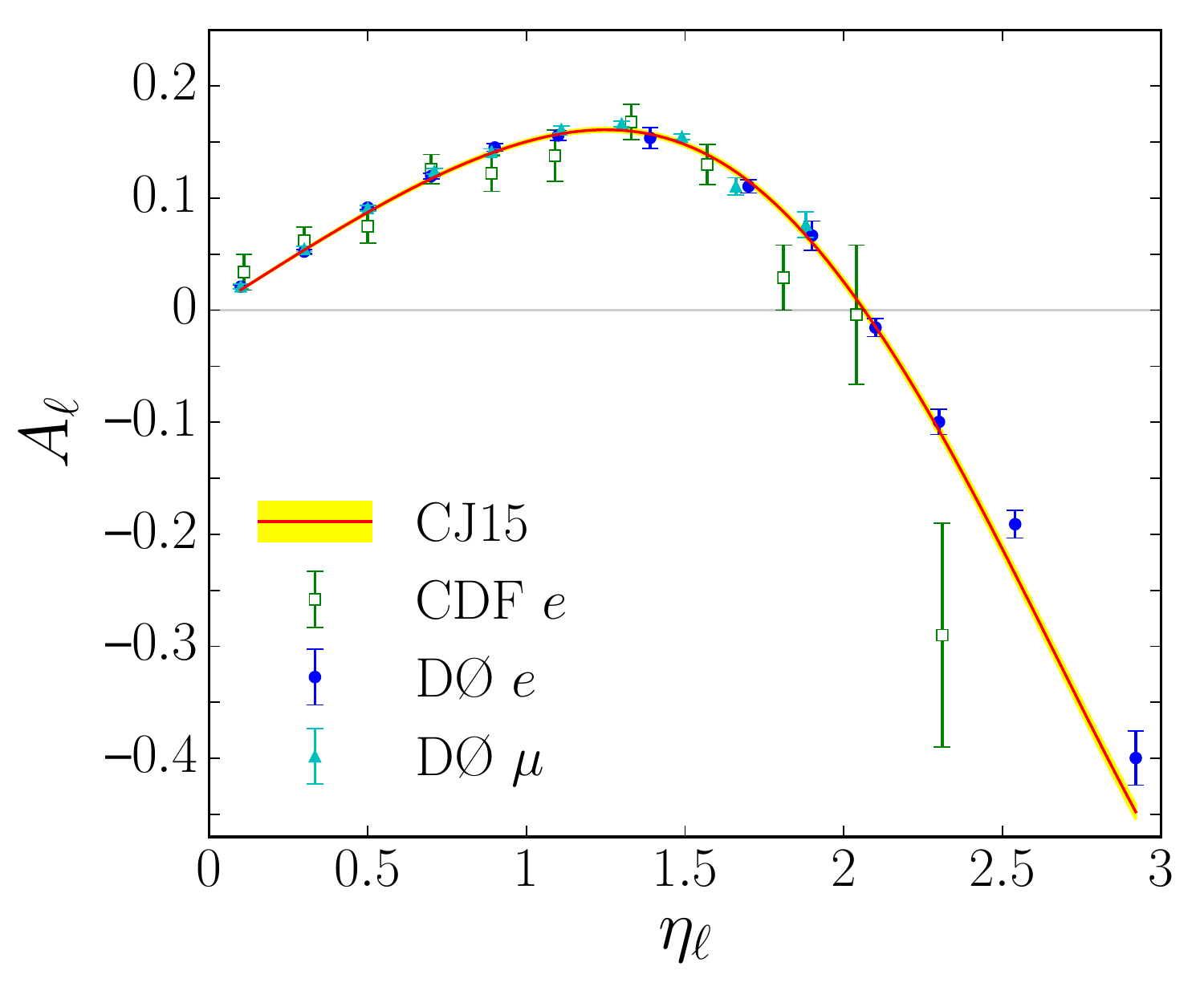}
\caption{Lepton charge asymmetry $A_{\ell}$ from
	$p\bar p \to W X \to \ell\, \nu X$ as a function of the
	lepton pseudorapidity $\eta_{\ell}$ from
	CDF electron (green open squares) \cite{CDF_e},
	D\O\ electron (blue circles) \cite{D0_e} and
	D\O\ muon (cyan triangles) \cite{D0_mu} data compared with
	the CJ15 fit with 90\% CL uncertainty (yellow band).}
\label{fig:Lasy}
\end{figure}

\begin{figure}[t]
\includegraphics[width=15cm]{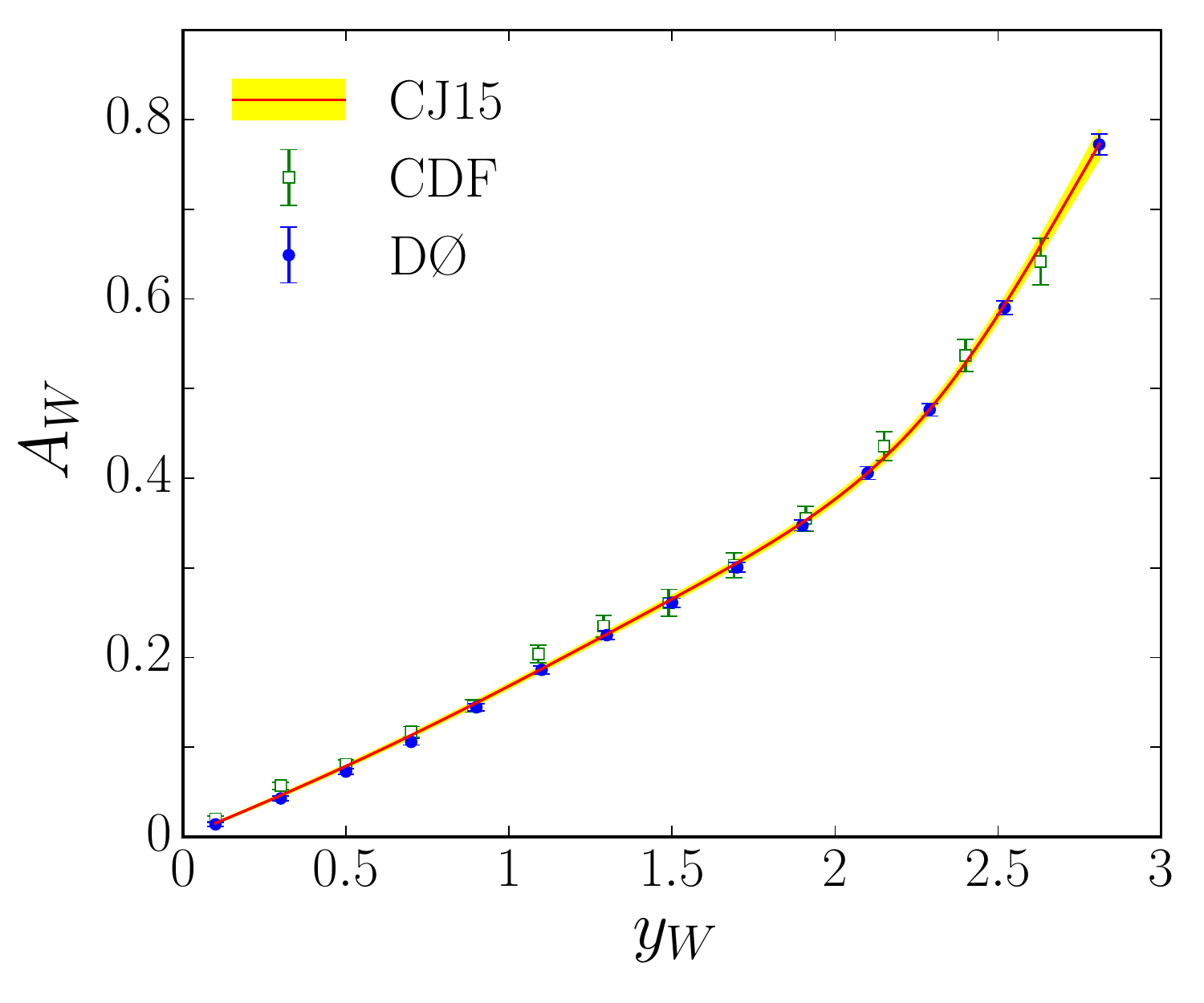}
\caption{$W$ boson charge asymmetry $A_W$ from $p\bar p \to W X$
	as a function of the $W$ boson rapidity $y_W$ for
	CDF (green open squares) \cite{CDF_W} and
	D\O\ (blue circles) \cite{D0_W} data compared with the
	CJ15 fit with 90\% CL uncertainty (yellow band).}
\label{fig:Wasy}
\end{figure}

Since our last analysis, the new data from the D\O\ collaboration
on charged lepton \cite{D0_mu, D0_e} and $W$ boson asymmetries
\cite{D0_W} that have become available have allowed significant
new constraints to be placed on the $d/u$ ratio at high $x$.
The new D\O\ electron and muon asymmetry data, together with earlier
data from CDF \cite{CDF_e}, are displayed in Fig.~\ref{fig:Lasy}
as a function of the lepton pseudorapidity $\eta_l$ and compared
with the CJ15 fit.
The extracted $W$ boson asymmetries, which are more directly related
to the shape of the PDFs and are not limited in their $x$ reach by the
lepton decay vertex smearing, are shown in Fig.~\ref{fig:Wasy} as a
function of the $W$ boson rapidity $y_W$.  The statistical errors
on the D\O\ data in particular are extremely small and place strong
constraints on the fit.  The earlier CDF electron and $W$ data
have larger errors and have more limited constraining power.
Compared with the range of nuclear corrections in CJ12, the
asymmetry data, and especially the new results from D\O, strongly
favor smaller nuclear corrections at large $x$, closer to those in
the CJ12min set.

\begin{figure}[t]
\includegraphics[width=15cm]{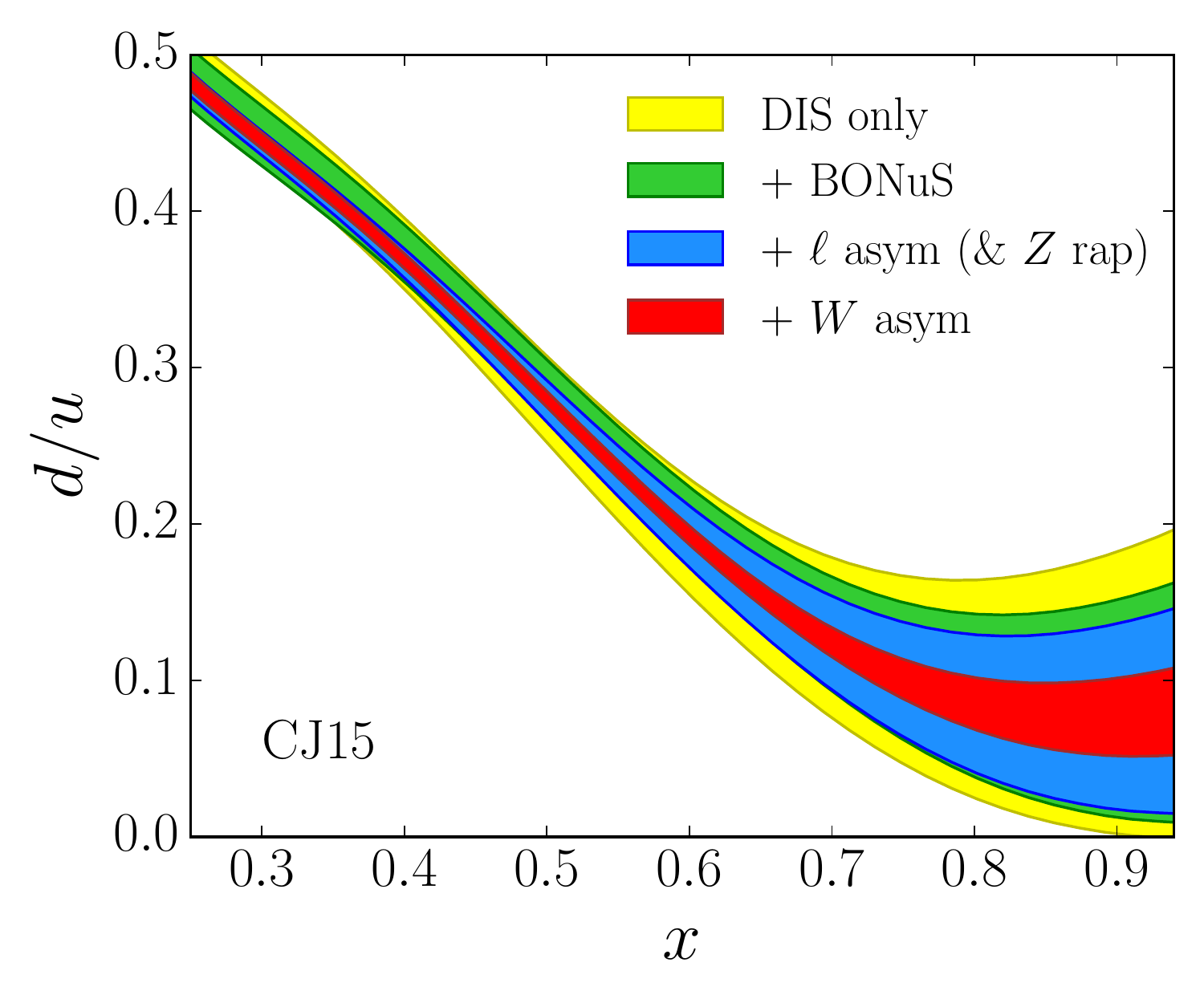}
\caption{Impact of various data sets on the $d/u$ ratio at
	$Q^2=10$~GeV$^2$. The 90\% CL uncertainty band is
	largest for the DIS only data (yellow band), and decreases
	with the successive addition of Jefferson Lab BONuS
	$F_2^n/F_2^d$ \cite{Tkachenko14} data (green band),
	lepton asymmetry \cite{CDF_e, D0_mu, D0_e}
	(and $Z$ rapidity \cite{CDFZ, D0Z}) data (blue band), and
	$W$ boson asymmetry data \cite{CDF_W, D0_W} (red band).}
\label{fig:du_data}
\end{figure}

The stronger constraints from the lepton and $W$ charge asymmetry
data lead to a significant reduction in the uncertainties on the $d/u$
ratio, particularly at large values of $x$.  This is illustrated in
Fig.~\ref{fig:du_data}, which demonstrates the shrinking of the $d/u$
uncertainty bands (which are shown here and in the remainder of this
section at the 90\% CL) with the successive addition of various data
sets.
Compared with the fit to DIS only data, in which the $d/u$ ratio has
very large uncertainties beyond $x \approx 0.4$, the addition of the
lepton asymmetries leads to a reduction in $d/u$ of more than a factor
of two at $x \lesssim 0.4$, with more limited impact at higher $x$
values due to the PDF smearing caused by the lepton decay vertex.
(Addition of $Z$ boson rapidity data \cite{CDFZ, D0Z} has only
modest impact on $d/u$.)
Subsequent inclusion of the $W$ asymmetries leads to a further
halving of the uncertainty at $x \approx 0.6-0.8$, while having
minimal effect on the errors at $x \lesssim 0.4$.

In fact, independent of the charge asymmetry data, a significant
reduction in the $d/u$ uncertainty at intermediate $x$ values is
already provided by the Jefferson Lab BONuS data on $F_2^n/F_2^d$
\cite{Baillie12, Tkachenko14}.  While the BONuS data have little or
no effect at $x \lesssim 0.3$, the reduction in the $d/u$ error at
$x \sim 0.5-0.6$ is almost as large as that from the lepton asymmetries.
(The BONuS data have a slight preference for stronger nuclear
corrections, in contrast to the lepton asymmetry data, although
the tension is not significant.)
Using all the available data from DIS and $W$ boson production,
the central value of the extrapolated $d/u$ ratio at $x=1$
is $\approx 0.1$ at the input scale $Q_0^2$.
The nuclear model dependence of the central values of the
$x \to 1$ limit of $d/u$ is relatively weak, ranging from 0.08
for the WJC-1 wave function to 0.12 for the CD-Bonn model.
For our best fit we obtain the extrapolated value
\begin{align}
d/u\ \xrightarrow[x \to 1]\ \ 0.09 \pm 0.03
\end{align}
at the 90\% CL, which represents a factor $\approx 2$ reduction
in the central value compared with the CJ12 result \cite{CJ12}.

\begin{figure}[t]
\includegraphics[width=15cm]{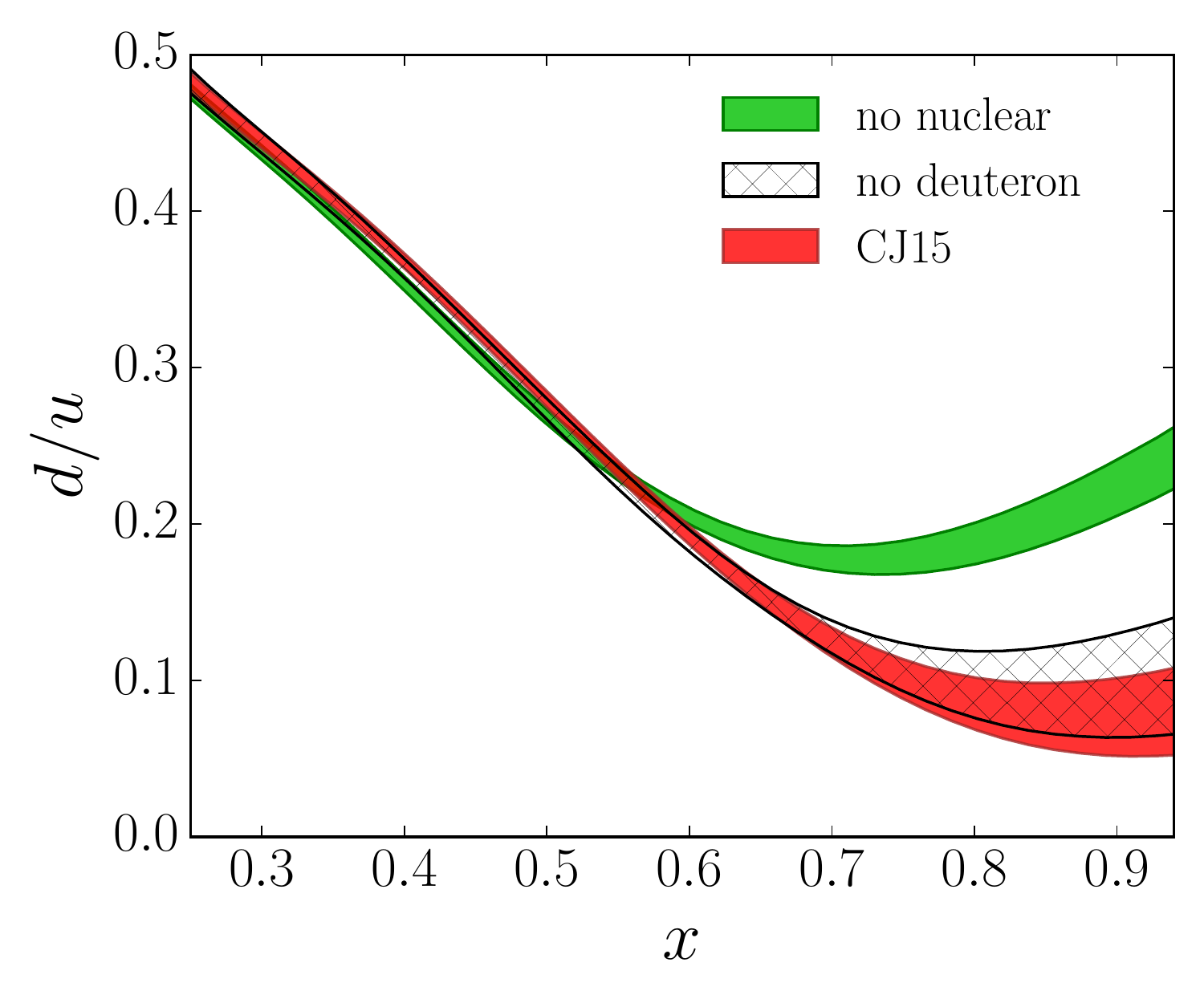}
\caption{Impact on the CJ15 $d/u$ ratio at $Q^2=10$~GeV$^2$
	(red band) of removing the deuterium nuclear corrections
	(green band), and omitting the deuterium data
	(cross-hatched band).}
\label{fig:du_nuc}
\end{figure}

While the new charge asymmetry and BONuS $F_2^n/F_2^d$ measurements
provide important constraints on the $d/u$ ratio, the existing
inclusive deuteron DIS data still play an important role in global
analyses, as does the proper treatment of the nuclear corrections.
If one were to fit $F_2^d$ data without accounting for nuclear
effects (assuming $F_2^d = F_2^p + F_2^n$), the resulting $d/u$
ratio would be strongly overestimated beyond $x = 0.6$, where the
$F_2^d/F_2^N$ ratio begins to deviate significantly from unity
(see Fig.~\ref{fig:F2dN}).
This is illustrated in Fig.~\ref{fig:du_nuc}, where the CJ15
$d/u$ ratio is compared with the fit without nuclear corrections.
This behavior can be understood from the shape of the $F_2^d/F_2^N$
ratio Fig.~\ref{fig:F2dN} at large $x$, where the effect of the
nuclear corrections is to increase the ratio above unity for
$x \gtrsim 0.6$.  Since $F_2^d$ and $F_2^p$ are fixed inputs,
a larger $F_2^d/F_2^N$ is generated by a smaller neutron $F_2^n$
and hence a smaller $d/u$ ratio.  For example, the effect of the
nuclear corrections is to shift the $d/u$ ratio at $x = 0.8$
from the range $\approx 0.1 - 0.3$ to $\approx 0 - 0.2$ once
the smearing and off-shell effects are included.
Removing the deuterium data altogether increases the overall
uncertainty band for $x \gtrsim 0.7$.  The deuteron data also
reduce the $d/u$ uncertainties slightly at smaller values of
$x \lesssim 0.2$ (see below).

Effects on large-$x$ PDFs from nuclear corrections have also been
investigated by several other groups in recent years \cite{MMHT14,
JR14, MMSTWW13, ABKM09, ABM11} and it is instructive to compare
the CJ15 results on the $d/u$ ratio with those analyses.
The MMHT14 fit \cite{MMHT14} uses a purely phenomenological, 
$Q^2$-independent nuclear correction for the combined effects of
nuclear smearing and off-shell corrections, in contrast to our
approach in which the (poorly understood) off-shell correction is
fitted, but the (better known) deuteron wave function correction
is computed, and finite-$Q^2$ effects are taken into account.
Interestingly, the phenomenological MMHT14 $F_2^d/F_2^N$ ratio
has a qualitatively similar shape to that found in our more
microscopic estimate, which offers an important cross check
of our formalism.  For $x \lesssim 0.7$ the MMHT14 $d/u$
uncertainty is comparable to that in CJ15, although for
$x \gtrsim 0.8$ the uncertainty diverges rapidly due to the
adoption of a stiffer $d$-quark parametrization, which only
allows the $d/u$ ratio to approach zero or infinity as $x \to 1$.

The JR14 analysis \cite{JR14} uses similar smearing functions to
those used in our fit, but does not include nucleon off-shell
corrections.  Furthermore, it uses the $\Delta\chi^2 = 1$
criterion for the 1$\sigma$ CL, based on statistical
considerations alone, introducing additional systematic
uncertainties through the dependence of the fit on the input scale.
The resulting uncertainty on $d/u$ is larger than that in CJ15 in
the intermediate-$x$ region, which may reflect the absence of
the recent $W$ and lepton asymmetry data in the JR14 fit.
The range of $d/u$ values extrapolated to $x = 1$ is similar
to the CJ15 band within errors, although the form of the JR14
parametrization forces $d/u \to 0$ at $x=1$.

\begin{figure}[t]
\includegraphics[width=15cm]{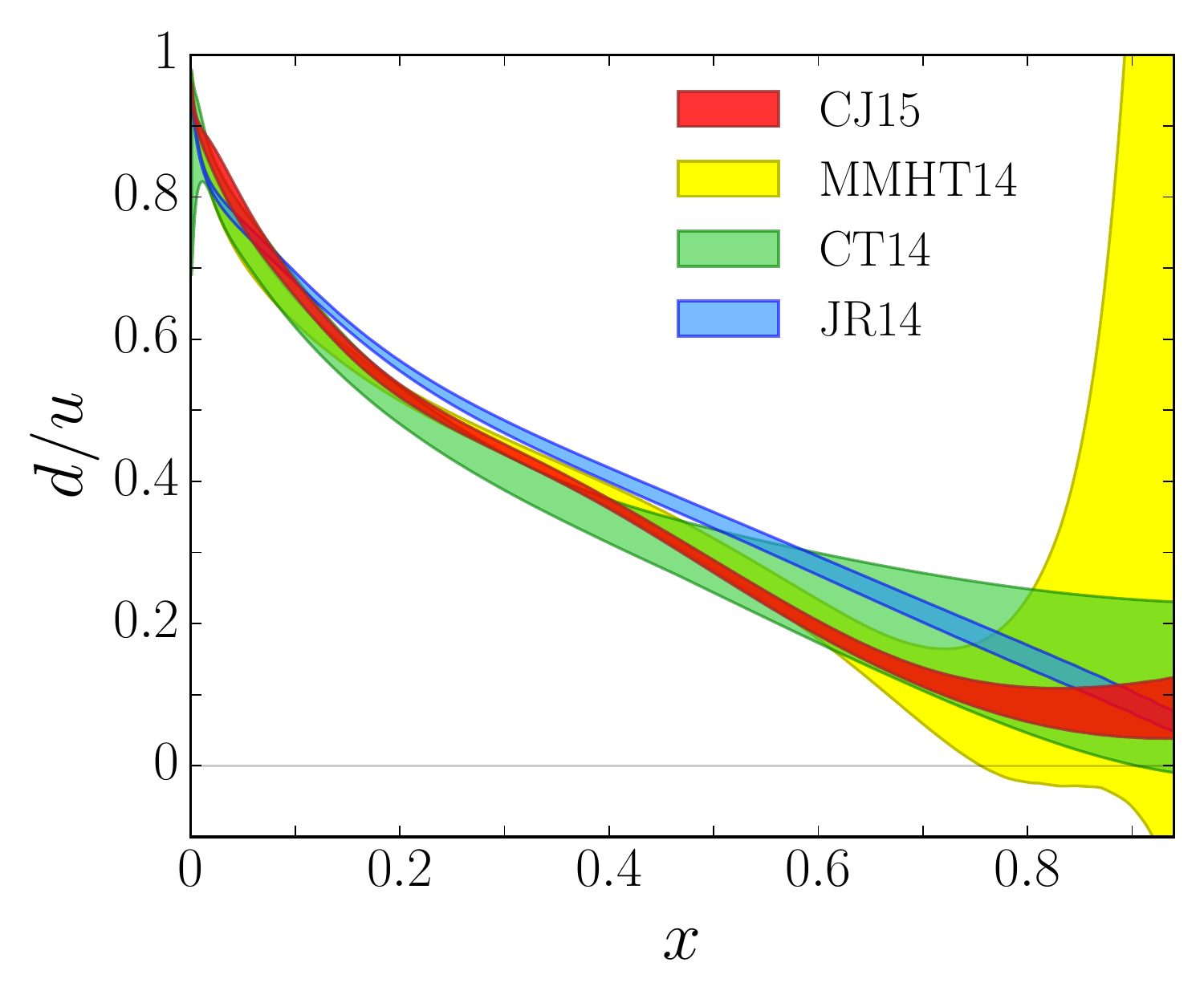}
\caption{Comparison of the $d/u$ ratio at $Q^2=10$~GeV$^2$ for
	different PDF parametrizations:
	CJ15 (red band),
	MMHT14 \cite{MMHT14} (yellow band, 68\% CL),
	CT14 \cite{CT14} (green band), and
	JR14 \cite{JR14}
	(blue band, scaled by a factor 1.645 for the 90\% CL).}
\label{fig:du_pdfs}
\end{figure}

The CJ15 uncertainty band in Fig.~\ref{fig:du_pdfs} is also similar
to that found in the CT14 global analysis \cite{CT14}, which does not
apply any nuclear corrections to deuterium data, on the basis of the
somewhat higher $W^2$ cuts utilized.  The CT14 analysis uses a
parametrization based on Bernstein polynomials multiplying a
common factor $x^{a_1} (1-x)^{a_2}$, and fixes the exponents $a_2$
to be the same for the $u$- and $d$-quark PDFs, thereby allowing
finite values of the $d/u$ ratio in the $x \to 1$ limit.
The results of the two analyses largely overlap over much of the
$x$ range, with the CT14 distributions being slightly above the
CJ15 error band at $x \gtrsim 0.6$.  This is reminiscent of the
higher $d/u$ ratio observed in Fig.~\ref{fig:du_nuc} when the
nuclear corrections are switched off.

\begin{figure}[t]
\includegraphics[width=15cm]{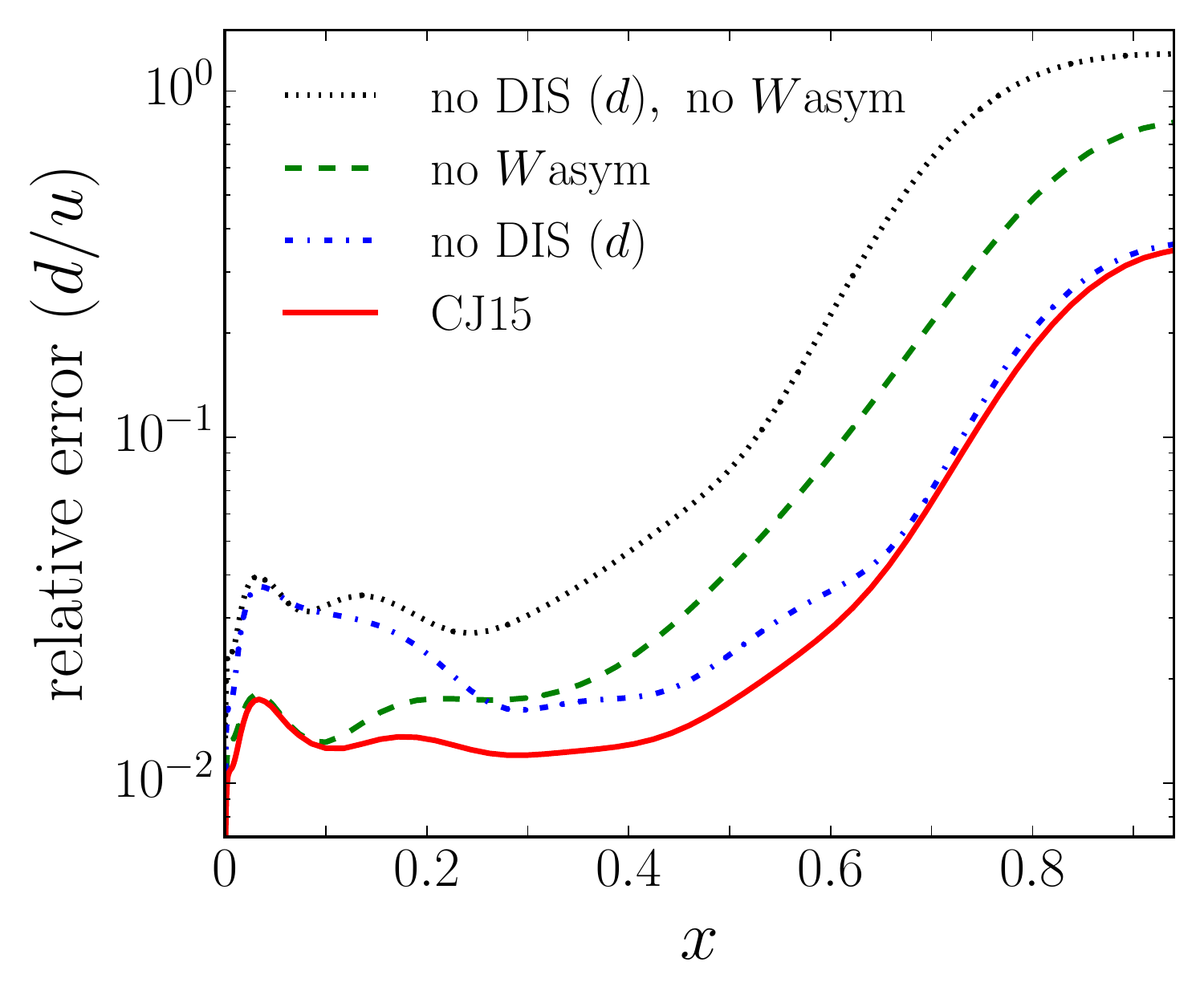}
\caption{Relative error on the $d/u$ PDF ratio versus $x$ at
	$Q^2=10$~GeV$^2$ from the CJ15 fit
	(90\% CL, solid red curve)
	compared with the uncertainties obtained in fits
	excluding deuteron DIS data (dot-dashed blue curve) or
	$W$ asymmetry data (dashed green curve), as well as
	excluding both (dotted black curve).}
\label{fig:du_err}
\end{figure}

Finally, in Fig.~\ref{fig:du_err} we show the $d/u$ uncertainty
from the CJ15 fit compared with the uncertainties obtained in fits
excluding DIS deuteron or $W$ asymmetry data.  The $W$ asymmetry
data, which are statistically dominated by the D\O\ results,
provide the main constraint on the $d/u$ ratio at $x \gtrsim 0.3$.
At $x \lesssim 0.3$, where the statistical power of the
reconstructed $W$ asymmetry data becomes limited, the global
deuteron DIS data play a vital role in reducing the uncertainty
in the $d/u$ ratio by more than 50\%.  
At $x\gtrsim 0.6$, the statistical power of the DIS data is
utilized instead to fit the off-shell function $\delta f^N$.
The combination of these two observables provides a good
illustration of the complementarity of different data sets in
global fits in constraining PDFs over extended regions of $x$.

\newpage
\section{Conclusion}
\label{sec:conclusion}

We have presented here results of the CJ15 global NLO analysis of
parton distributions, taking into account the latest developments in
theory and the availability of new data.  Focusing particularly, but
not exclusively, on the large-$x$ region, the new analysis features
a more comprehensive treatment of nuclear corrections to deuterium
data, as well as a more flexible parametrization of the SU(2) light
antiquark asymmetry, and an improved treatment of heavy quarks.
In contrast to the earlier CJ12 fit \cite{CJ12}, which used
physically motivated models for the nucleon off-shell corrections,
the present analysis allows the magnitude and shape of the off-shell
effects to be phenomenologically constrained directly from data.

Along with the expanded set of proton and deuteron DIS data afforded 
by our less restrictive kinematic cuts $Q^2 > (1.3~{\rm GeV})^2$
and $W^2 > 3$~GeV$^2$, we also include new results from the BONuS
experiment at Jefferson Lab \cite{Baillie12, Tkachenko14}, which
provide the first determination of the neutron structure function
essentially free of nuclear correction uncertainties.
The greatest impact on the fits, however, comes from the new D\O\
$W$ asymmetry data at large rapidity \cite{D0_W}, which because
of their high precision and kinematic reach are able to place
significant constraints on PDFs at high $x$.
In particular, while the previous CJ12 analysis provided three sets
of PDFs corresponding to a range of different deuterium and off-shell
models, the new $W$ asymmetry data strongly favor models with
smaller nuclear corrections, closer to the ``CJ12min'' PDF set
\cite{CJ12}.  Within the parametrization of the nucleon off-shell
corrections adopted here, our analysis has a slight preference
for deuteron wave functions with softer momentum distributions,
but essentially indistinguishable fits can be obtained with each
of the deuteron models considered.

Our approach to the nuclear corrections is similar in spirit to the
phenomenological analysis of Ref.~\cite{KP06}, which makes use of
DIS data on a wide range of nuclear targets and finds the ratio
$F_2^d/F_2^N$ to have a universal shape similar to that for
$F_2^A/F_2^d$ for heavy nuclei.  From the proton and deuterium
data alone, however, we find no evidence for an enhancement of
$F_2^d/F_2^N$ in the vicinity of $x \approx 0.1$.  The only way to
definitively resolve this question may be with data on the free
neutron structure function that are not subject to assumptions
about nuclear corrections in deuterium.
The phenomenological approach of fitting the nuclear effects directly
was also utilized in Refs.~\cite{MMHT14, MMSTWW13}, who parametrized
the entire nuclear correction by a function that mimics both the
effects of the smearing and the nucleon off-shell correction.
Since the nuclear physics of the deuteron at long distances is
relatively well understood, our philosophy is to include in the
theoretical description the effects that can be computed reliably,
and parametrize those that are more strongly model dependent.

As anticipated in Refs.~\cite{CJ12, Accardi13} and elaborated in
Ref.~\cite{Accardi16}, the new precision measurements of observables
that are sensitive to the $d$-quark PDF, but less sensitive to nuclear
corrections, are seen to play an important role in allowing global QCD
fits to constrain models of nuclear corrections in the deuteron.
In particular, a simultaneous fit of the new $W$ charge asymmetries
\cite{D0_W} and the SLAC deuteron DIS structure functions \cite{SLAC}
is only possible when nuclear corrections are taken into account.
The interplay of these two data sets within the CJ15 fit has provided
the first determination of nucleon off-shell effects in quark
distributions in the deuteron within a global QCD context.
At the same time, the $d/u$ ratio has seen a significant reduction
in its uncertainty at $x \gtrsim 0.5$, with an extrapolated central
value $\approx 0.1$ at $x \to 1$, or about half of that found in the
CJ12 fit \cite{CJ12}.
As discussed in Refs.~\cite{Brady12, PDFLHC16}, a precise determination
of the $d$-quark PDF at large $x$ is vital for searches for physics
beyond the standard model at the LHC at the edges of kinematics,
such as at large rapidities in heavy weak-boson production, or more
generally in large invariant mass observables.

The uncertainty in the $d/u$ ratio is expected to be further reduced
once new data from experiments at the energy-upgraded Jefferson Lab
facility become available \cite{MARATHON, BONUS12, SOLID}, that will
probe PDFs up to $x \sim 0.85$ at DIS kinematics.
The first of these, involving the simultaneous measurement of
inclusive DIS cross section from $^3$He and $^3$H \cite{MARATHON},
in which the nuclear corrections are expected to mostly cancel
\cite{Afnan00, Pace01, SSS02}, is scheduled to begin data taking
in Fall 2016.  The current analysis provides a timely benchmark against
which the upcoming experimental results can be calibrated.

\newpage
\acknowledgments

We thank E.~Christy, C.~Keppel, P.~Monaghan and S.~Malace for
their collaboration and assistance with the experimental data sets,
and S.~Kulagin and R.~Petti for helpful discussions.
This work was supported by the DOE contract No.~DE-AC05-06OR23177,
under which Jefferson Science Associates, LLC operates Jefferson Lab.
The work of J.F.O. and A.A. was supported in part by DOE contracts
No.~DE-FG02-97ER41922 and No.~DE-SC0008791, respectively.


\clearpage

\clearpage

\clearpage

\clearpage

\clearpage

\clearpage

\clearpage

\clearpage

\clearpage

\clearpage

\clearpage

\clearpage
  
\clearpage

\clearpage

\clearpage

\clearpage

\clearpage

\newpage
\begin{table}[t]
\caption{Data sets used in the CJ15 global analysis, with the
	corresponding number of data points and $\chi^2$ values
	for each set.  The main CJ15 NLO fit (in boldface),
	which uses the AV18 deuteron wave function and off-shell
	parametrization in Eq.~(\ref{eq:delffit}), is compared
	with an LO fit and NLO fits with the OCS off-shell model,
	no nuclear corrections, and no nuclear corrections or
	D\O\ $W$ asymmtetry data.}
\centering
{\scriptsize 
\newcolumntype{C}[1]{>{\hsize=#1\centering\arraybackslash}X}
\centering
\begin{tabularx}{0.92\linewidth}{llc*{4}{C{1.3cm}}C{1.75cm}}  \hline
Observable
  &  Experiment & \# points         & \multicolumn{5}{c}{\ \ \ \ \ $\chi^2$} \\
  &  		& 		    & LO & NLO & NLO  & NLO       & NLO          \\
  & 		&		    &    &     &(OCS) & (no nucl) & (no nucl/D0) \\ \hline
DIS $F_2$
  & BCDMS $(p)$		\cite{BCDMS}	& 351 &	 426 & {\bf 438} & 436	&  440 &  427 \\
  & BCDMS $(d)$		\cite{BCDMS}	& 254 &	 292 & {\bf 292} & 289	&  301 &  301 \\
  & SLAC  $(p)$		\cite{SLAC}	& 564 &	 480 & {\bf 434} & 435	&  441 &  440 \\
  & SLAC  $(d)$		\cite{SLAC}	& 582 &	 415 & {\bf 376} & 380	&  507 &  466 \\
  & NMC	  $(p)$		\cite{NMCp}	& 275 &	 416 & {\bf 405} & 404	&  405 &  403 \\
  & NMC	  $(d/p)$	\cite{NMCdop}	& 189 &	 181 & {\bf 172} & 173	&  174 &  173 \\
  & HERMES $(p)$	\cite{HERMES}	&  37 &	  57 & {\bf  42} &  43	&   44 &   44 \\
  & HERMES $(d)$	\cite{HERMES}	&  37 &	  52 & {\bf  37} &  38	&   36 &   37 \\
  & Jefferson Lab $(p)$	\cite{Malace}	& 136 &	 172 & {\bf 166} & 167	&  177 &  166 \\
  & Jefferson Lab $(d)$	\cite{Malace}	& 136 &	 131 & {\bf 123} & 124	&  126 &  130 \\
DIS $F_2$ tagged							     
  & Jefferson Lab $(n/d)$ \cite{Tkachenko14} & 191 &	 216 & {\bf 214} & 213	&  219 &  219 \\
DIS $\sigma$								     
  & HERA (NC $e^-p$)	\cite{HERA2}	& 159 &	 315 & {\bf 241} & 240	&  247 &  244 \\
  & HERA (NC $e^+p$ 1)	\cite{HERA2}	& 402 &	 952 & {\bf 580} & 579	&  588 &  585 \\
  & HERA (NC $e^+p$ 2)	\cite{HERA2}	&  75 &	 177 & {\bf  94} &  94	&   94 &   93 \\
  & HERA (NC $e^+p$ 3)	\cite{HERA2}	& 259 &	 311 & {\bf 249} & 249	&  248 &  248 \\
  & HERA (NC $e^+p$ 4)	\cite{HERA2}	& 209 &	 352 & {\bf 228} & 228	&  228 &  228 \\
  & HERA (CC $e^-p$)	\cite{HERA2}	&  42 &	  42 & {\bf  48} &  48	&   45 &   49 \\
  & HERA (CC $e^+p$)	\cite{HERA2}	&  39 &	  53 & {\bf  50} &  50	&   51 &   51 \\
Drell-Yan								     
  & E866 $(pp)$		\cite{E866}	& 121 &	 148 & {\bf 139} & 139	&  145 &  143 \\
  & E866 $(pd)$		\cite{E866}	& 129 &	 202 & {\bf 145} & 143	&  158 &  157 \\
$W/$charge asymmetry							     
  & CDF	 ($e$)		\cite{CDF_e}	&  11 &	  11 & {\bf  12} &  12	&   13 &   14 \\
  & D\O\ ($\mu$)	\cite{D0_mu}	&  10 &	  18 & {\bf  20} &  19	&   29 &   28 \\
  & D\O\ ($e$)		\cite{D0_e}	&  13 &	  49 & {\bf  29} &  29	&   14 &   14 \\
  & CDF	 ($W$)		\cite{CDF_W}	&  13 &	  16 & {\bf  16} &  16	&   14 &   14 \\
  & D\O\ ($W$)		\cite{D0_W}	&  14 &	  35 & {\bf  14} &  15	&   82 &   --- \\
$Z$ rapidity								     
  & CDF	 ($Z$)		\cite{CDFZ}	&  28 &	 108 & {\bf  27} &  27	&   26 &   26 \\
  & D\O\ ($Z$)		\cite{D0Z}	&  28 &	  26 & {\bf  16} &  16	&   16 &   16 \\
jet									     
  & CDF	 (run 2)	\cite{CDFjet2}	&  72 &	  29 & {\bf  15} &  15	&   23 &   25 \\
  & D\O\ (run 2)	\cite{D0jet2}	& 110 &	  90 & {\bf  21} &  21	&   14 &   14 \\
$\gamma$+jet								     
  & D\O\ 1		\cite{D0gamjet} &  16 &	  16 & {\bf   7} &   7	&    7 &    7 \\
  & D\O\ 2		\cite{D0gamjet} &  16 &	  34 & {\bf  16} &  16	&   17 &   17 \\
  & D\O\ 3		\cite{D0gamjet} &  12 &	  35 & {\bf  25} &  25	&   24 &   25 \\
  & D\O\ 4		\cite{D0gamjet} &  12 &	  79 & {\bf  13} &  13	&   13 &   13 \\ \hline
total		   &                   & 4542 & 5935 & {\bf 4700}& 4702 & 4964 & 4817 \\
total + norm	   &                   &      & 6058 & {\bf 4708}& 4710 & 4972 & 4826 \\ \hline
$\chi^2$/datum     &                   &      & 1.33 & {\bf 1.04}& 1.04 & 1.09 & 1.07 \\ \hline\\
\end{tabularx}
}
\label{tab:chi2}
\end{table}

\begin{table}[t]
\begin{center}
\caption{Leading twist parameter values and the 1$\sigma$ uncertainties
	for the	$u_v$, $d_v$, $\bar d+\bar u$, $\bar d/\bar u$ and $g$
	PDFs [Eqs.~(\ref{eq:param}), (\ref{eq:dboub})] from the CJ15 NLO
        analysis at the input scale $Q_0^2$.  Parameters without
	errors have been fixed.  For the $d_v$ PDF, the large-$x$
	parameters [Eq.~(\ref{eq:du})] are given by
	$b = (3.6005 \pm 0.66324) \times 10^{-3}$ with $c=2$.
	For the strange to nonstrange sea quark PDF ratio
	[Eq.~(\ref{eq:kappa})], we take $\kappa=0.4$.
	(The parameter values are given to 5 significant
	figures	to avoid rounding errors.)\\}
{\scriptsize
\begin{tabular}{c|ccccc}\hline
parameter	& $u_v$
		& $d_v$
		& $\bar d+\bar u$
		& $\bar d/\bar u$
		& $g$				\\ \hline
$a_0$		& 2.4067
		& 24.684
		& $0.14658 \pm 0.0050348$
		& 35712
		& 45.542			\\
$a_1$		&~$0.61537 \pm 0.019856$~
		& $1.1595  \pm 0.033533$
		& $-0.20775\pm 0.0037551$
		&~$4.0249  \pm 0.07407$~
		&~$0.60307 \pm 0.031164$~	\\
$a_2$		& $3.5433  \pm 0.012414$
		& $6.5514  \pm 0.15936$
		& $8.3286  \pm 0.19114$
		& $20.154  \pm 0.87862$
		& $6.4812  \pm 0.96748$		\\
$a_3$		& 0
		&~$-3.5030 \pm 0.086332$~
		& 0
		& 17
		& $-3.3064 \pm 0.13418$		\\
$a_4$		& $3.4609  \pm 0.42903$
		& $4.6787  \pm 0.14209$
		& $14.606  \pm 1.2151$
		& $51.156  \pm 10.239$
		& $3.1721  \pm 0.31376$		\\ \hline
\end{tabular}
}
\label{tab:LTparams}
\end{center}
\end{table}

\begin{table}[h]
\begin{center}
\caption{Parameter values and 1$\sigma$ uncertainties
	for the nucleon off-shell [Eq.~(\ref{eq:delffit})]
	and higher twist [Eq.~(\ref{eq:C_ht})] corrections to $F_2$
	from the CJ15 NLO analysis at the input scale $Q_0^2$.
	The off-shell parameters are fitted using the
	AV18 deuteron wave function.
	Parameters without errors have been fixed.
	(The parameter values are given to 5 significant
	figures to avoid rounding errors.)
        The covariance matrix is provided for all the fitted
	off-shell and higher twist parameters.\\}
{\scriptsize
\begin{tabular}{c|c|ccc}\hline
parameter & value                               &  \multicolumn{3}{c}{covariance matrix} \\ \hline
$C$	  & $-3.6735 \pm 1.5278$		& \ \ 1.000 &  $-$0.173 & ---  \\
$x_0$	  & $(5.7717 \pm 1.4842) \times 10^{-2}$&  $-$0.173 & \ \ 1.000 & ---  \\
$x_1$     & 0.36419                             &  ---      &  ---      & ---  \\ \hline
$h_0$     & $-3.2874 \pm 0.26061$		& \ \ 1.000 &  $-$0.812 &  $-$0.497 \\
$h_1$     & $ 1.9274 \pm 0.10524$		& $-$0.812  & \ \ 1.000 & \ \ 0.119 \\
$h_2$     & $-2.0701 \pm 0.019888$              & $-$0.497  & \ \ 0.119 & \ \ 1.000 \\ \hline
\end{tabular}
}
\label{tab:other_params}
\end{center}
\end{table}


\begin{thebibliography}{99}

\bibitem{JMO13}
P.~Jimenez-Delgado, W.~Melnitchouk and J.~F.~Owens,
J. Phys. G: Nucl. Part. Phys. {\bf 40}, 093102 (2013).

\bibitem{Blumlein13}
J.~Bl\"umlein,
Prog. Part. Nucl. Phys. {\bf 69}, 28 (2013).

\bibitem{ForteWatt13}
S.~Forte and G.~Watt,
Ann. Rev. of Nucl. Part. Sci. {\bf 63}, (2013).

\bibitem{Cooper-Sarkar15}
A.~M.~Cooper-Sarkar,
arXiv:1507.03849 [hep-ph].

\bibitem{JLab11}
M.~E.~Christy and W.~Melnitchouk,
J. Phys.: Conf. Ser. {\bf 299}, 012004 (2011).

\bibitem{MMHT14}
L.~A.~Harland-Lang, A.~D.~Martin, P.~Motylinski and R.~S.~Thorne,
Eur. Phys. J. C {\bf 75}, 204 (2015).

\bibitem{CT14} 
S.~Dulat {\it et al.},
arXiv:1506.07443 [hep-ph].

\bibitem{NNPDF3.0}
R.~D.~Ball {\it et al.},
J. High Energy Phys. {\bf 04} (2015) 040.

\bibitem{HERAPDF15}
V.~Radescu,
in Proceedings of 35th International Conference of High Energy Physics
(ICHEP2010), Paris, France (2010),
arXiv:1308.0374.

\bibitem{JR14}
P.~Jimenez-Delgado and E.~Reya,
Phys. Rev. D {\bf 89}, 074049 (2014).

\bibitem{ABMP15}
S.~Alekhin, J.~Bl\"uemlein, S.~Moch and R.~Placakyte,
arXiv:1508.07923 [hep-ph].

\bibitem{CJ10}
A.~Accardi, M.~E.~Christy, C.~E.~Keppel, P.~Monaghan, W.~Melnitchouk,
J.~G.~Morfin and J.~F.~Owens,
Phys. Rev. D {\bf 81}, 034016 (2010).

\bibitem{CJ11}
A.~Accardi, W.~Melnitchouk, J.~F.~Owens, M.~E.~Christy, C.~E.~Keppel,
L.~Zhu and J.~G.~Morfin,
Phys. Rev. D {\bf 84}, 014008 (2011).

\bibitem{CJ12}
J.~F.~Owens, A.~Accardi and W.~Melnitchouk,
Phys. Rev. D {\bf 87}, 094012 (2013).

\bibitem{CJweb}
The CTEQ-Jefferson Lab (CJ) Collaboration website,
\url{http://www.jlab.org/cj}.

\bibitem{MT96}
W.~Melnitchouk and A.~W.~Thomas,
Phys. Lett. B {\bf 377}, 11 (1996).

\bibitem{D0_mu} 
V.~M.~Abazov {\it et al.},
Phys. Rev. D {\bf 88}, 091102 (2013).

\bibitem{D0_e} 
V.~M.~Abazov {\it et al.},
Phys. Rev. D {\bf 91}, 032007 (2015).

\bibitem{D0_W} 
V.~M.~Abazov {\it et al.},
Phys. Rev. Lett. {\bf 112}, 151803 (2014);
{\it ibid} {\bf 114}, 049901 (2015).

\bibitem{Baillie12}
N.~Baillie {\it et al.},
Phys. Rev. Lett. {\bf 108}, 142001 (2012).

\bibitem{Tkachenko14}
S.~Tkachenko {\it et al.},
Phys. Rev. C {\bf 89}, 045206 (2014).

\bibitem{S-ACOT}
M.~Kr\"amer, F.~I.~Olness and D.~E.~Soper,
Phys. Rev. D {\bf 62}, 096007 (2000).

\bibitem{Accardi13} 
A.~Accardi,
Mod. Phys. Lett. A {\bf 28}, 330032 (2013).

\bibitem{FJ75}
G.~R.~Farrar and D.~R.~Jackson,
Phys. Rev. Lett. {\bf 35}, 1416 (1975).

\bibitem{Close73}
F.~E.~Close,
Phys. Lett. B {\bf 43}, 422 (1973).

\bibitem{Close03}
F.~E.~Close and W.~Melnitchouk
Phys. Rev. C {\bf 68}, 035210 (2003).

\bibitem{HR10}
R.~J.~Holt and C.~D.~Roberts,
Rev. Mod. Phys. {\bf 82}, 2991 (2010).

\bibitem{Roberts13} 
C.~D.~Roberts, R.~J.~Holt and S.~M.~Schmidt,
Phys. Lett. B {\bf 727}, 249 (2013).

\bibitem{E866}  
E.~A.~Hawker {\it et al.},
Phys. Rev. Lett. {\bf 80}, 3715 (1998);
%
J.~Webb,
Ph.D. Thesis, New Mexico State University (2002),
arXiv:hep-ex/0301031.

\bibitem{E866rat}
R.~S.~Towell {\it et al.},
Phys. Rev. D {\bf 64}, 052002 (2001).

\bibitem{SeaQuest}
Fermilab E906 experiment (SeaQuest),
{\it Drell-Yan Measurements of Nucleon and Nuclear Structure
with the Fermilab Main Injector},
D.~F.~Geesaman and P.~E.~Reimer, spokespersons;
{\tt http://www.phy.anl.gov/mep/SeaQuest/index.html}.

\bibitem{Accardi09}
A.~Accardi, F.~Arleo, W.~K.~Brooks, D.~D'Enterria and V.~Muccifora,
Riv. Nuovo Cim. {\bf 32}, 439 (2010).

\bibitem{Majumder11} 
A.~Majumder and M.~Van~Leeuwen,
Prog. Part. Nucl. Phys. A {\bf 66}, 41 (2011).

\bibitem{Eskola09} 
K.~J.~Eskola, H.~Paukkunen and C.~A.~Salgado,
J. High Energy Phys. {\bf 04} (2009) 065.

\bibitem{deFlorian11} 
D.~de~Florian, R.~Sassot, P.~Zurita and M.~Stratmann,
Phys. Rev. D {\bf 85}, 074028 (2012).

\bibitem{Kovarik15} 
K.~Kovarik {\it et al.},
arXiv:1509.00792 [hep-ph].

\bibitem{ATLAS-s}
G.~Aad {\it et al.},
Phys. Rev. Lett. {\bf 109}, 012001 (2012).

\bibitem{Alekhin-s}
S.~Alekhin, J.~Bl\"umlein, L.~Caminadac, K.~Lipka, K.~Lohwasser,
S.~Moch, R.~Petti and R.~Placakyte,
Phys. Rev. D {\bf 91}, 094002 (2015).

\bibitem{Catani04}
S.~Catani, D.~de Florian, G.~Rodrigo and W.~Vogelsang,
Phys. Rev. Lett. {\bf 93}, 152003 (2004).

\bibitem{Signal87}
A.~I.~Signal and A.~W.~Thomas,
Phys. Lett. B {\bf 191}, 205 (1987).

\bibitem{Zeller02}
G.~P.~Zeller {\it et al.},
Phys. Rev. D {\bf 65}, 111103 (2002)
[Phys. Rev. D {\bf 67}, 119902 (2003)].

\bibitem{ACOT}
F.~I.~Olness and W.-K.~Tung,
Nucl. Phys. {\bf B308}, 813 (1988);
%
M.~A.~Aivazis, F.~I.~Olness and W.-K.~Tung,
Phys. Rev. D {\bf 50}, 3085 (1994);
%
M.~A.~Aivazis, J.~C.~Collins, F.~I.~Olness and W.-K.~Tung,
Phys. Rev. D {\bf 50}, 3102 (1994).

\bibitem{GP76}
H.~Georgi and H.~D.~Politzer,
Phys. Rev. D {\bf 14}, 1829 (1976).

\bibitem{Schienbein08}
I.~Schienbein {\it et al.},
J. Phys. G {\bf 35}, 053101 (2008).

\bibitem{Brady11}
L.~T.~Brady, A.~Accardi, T.~J.~Hobbs and W.~Melnitchouk,
Phys. Rev. D {\bf 84}, 074008 (2011)
[Phys. Rev. D {\bf 85}, 039902 (2012)].

\bibitem{Greenberg71}
O.~W.~Greenberg and D.~Bhaumik,
Phys. Rev. D {\bf 4}, 2048 (1971).

\bibitem{Nachtmann73}
O.~Nachtmann,
Nucl. Phys. {\bf B63}, 237 (1973).

\bibitem{EFP76}
R.~K.~Ellis, R.~Petronzio and G.~Parisi,
Phys. Lett. B {\bf 64}, 97 (1976).

\bibitem{AHM09}
A.~Accardi, T.~Hobbs and W.~Melnitchouk,
J. High Energy Phys. {\bf 11} (2009) 084.

\bibitem{Guerrero15}
J.~V.~Guerrero, J.~J.~Ethier, A.~Accardi, S.~W.~Casper
and W.~Melnitchouk,
J. High Energy Phys. {\bf 09} (2015) 169.

\bibitem{AOT94} 
M.~A.~G.~Aivazis, F.~I.~Olness and W.~K.~Tung,
Phys. Rev. D {\bf 50}, 3085 (1994).

\bibitem{KR02}
S.~Kretzer and M.~H.~Reno,
Phys. Rev. D {\bf 66}, 113007 (2002).

\bibitem{AQ08} 
A.~Accardi and J.-W.~Qiu,
J. High Energy Phys. {\bf 07} (2008) 090.

\bibitem{SBMS12}
F.~M.~Steffens, M.~D.~Brown, W.~Melnitchouk and S.~Sanches,
Phys. Rev. C {\bf 86}, 065208 (2012).

\bibitem{Vir92}
M.~Virchaux and A.~Milsztajn,
Phys. Lett. B {\bf 274}, 221 (1992).

\bibitem{AKL03}
S.~I.~Alekhin, S.~A.~Kulagin and S.~Liuti,
Phys. Rev. D {\bf 69}, 114009 (2004).

\bibitem{BB08}
J.~Bl\"umlein and H. B\"ottcher,
Phys. Lett. B {\bf 662}, 336 (2008).

\bibitem{Blu12}
J.~Bl\"umlein,
Prog. Part. Nucl. Phys. {\bf 69}, 28 (2013).

\bibitem{Badelek92}
B.~Badelek and J.~Kwiecinski,
Nucl. Phys. {\bf B370}, 278 (1992).

\bibitem{MTshad}
W. Melnitchouk and A. W. Thomas
Phys. Rev. D {\bf 47}, 3783 (1993).

\bibitem{Kaptari91}
L.~P.~Kaptari and A.~Yu.~Umnikov,
Phys. Lett. B {\bf 272}, 359 (1991).

\bibitem{MSToff}
W.~Melnitchouk, A.~W.~Schreiber and A.~W.~Thomas,
Phys. Rev. D {\bf 49}, 1183 (1994).

\bibitem{KPW94}
S.~A.~Kulagin, G.~Piller and W.~Weise,
Phys. Rev. C {\bf 50}, 1154 (1994).

\bibitem{KMPW95}
S.~A.~Kulagin, W.~Melnitchouk, G.~Piller and W.~Weise,
Phys. Rev. C {\bf 52}, 932 (1995).

\bibitem{Ehlers14}
P.~J.~Ehlers, A.~Accardi, L.~T.~Brady and W.~Melnitchouk,
Phys. Rev. D {\bf 90}, 014010 (2014).

\bibitem{KP06}
S.~A.~Kulagin and R.~Petti,
Nucl. Phys. A {\bf 765}, 126 (2006).

\bibitem{KMK09}
Y.~Kahn, W.~Melnitchouk and S.~A.~Kulagin,
Phys. Rev. C {\bf 79}, 035205 (2009).

\bibitem{AV18}
R.~B.~Wiringa, V.~G.~J.~Stoks and R.~Schiavilla,
Phys. Rev. C {\bf 51}, 38 (1995).

\bibitem{CDBonn}
R.~Machleidt,
Phys. Rev. C {\bf 63}, 024001 (2001).

\bibitem{WJC}
F.~Gross and A.~Stadler,
Phys. Rev. C {\bf 78}, 014005 (2008);
{\it ibid.} C {\bf 82}, 034004 (2010).

\bibitem{ACHL09} 
J.~Arrington, F.~Coester, R.~J.~Holt and T.-S.~H.~Lee,
J. Phys. G {\bf 36}, 025005 (2009).

\bibitem{ARM12}
J.~Arrington, J.~G.~Rubin and W.~Melnitchouk,
Phys. Rev. Lett. {\bf 108}, 252001 (2012).

\bibitem{GL92} 
F.~Gross and S.~Liuti,
Phys. Rev. C {\bf 45}, 1374 (1992).

\bibitem{MSTplb}
W.~Melnitchouk, A.~W.~Schreiber and A.~W.~Thomas,
Phys. Lett. B {\bf 335}, 11 (1994).

\bibitem{MSS97}
W.~Melnitchouk, M.~Sargsian, and M.~Strikman,
Z. Phys. A {\bf 359}, 99 (1997).

\bibitem{Mineo04}
H.~Mineo, W.~Bentz, N.~Ishii, A.~W.~Thomas and K.~Yazaki,
Nucl. Phys. {\bf A735}, 482 (2004).

\bibitem{Cloet06}
I.~C.~Cloet, W.~Bentz and A.~W.~Thomas,
Phys. Lett. B {\bf 642}, 210 (2006).

\bibitem{EMC83}
J.~J.~Aubert {\it et al.},
Phys. Lett. B {\bf 123}, 275 (1983).

\bibitem{Geesaman95}
D.~F.~Geesaman, K.~Saito and A.~W.~Thomas,
Ann. Rev. Nucl. Part. Sci. {\bf 45}, 337 (1995).

\bibitem{MMSTWW13}
A.~D.~Martin, A.~J.~Th.~M.~Mathijssen, W.~J.~Stirling, R.~S.~Thorne, 
B.~J.~A.~Watt and G.~Watt,
Eur. Phys. J. C {\bf 73}, 2318 (2013).

\bibitem{BCDMS}
A.~C.~Benvenuti {\it et al.},
Phys. Lett. B {\bf 223}, 485 (1989);
{\it ibid.} B {\bf 236}, 592 (1989).

\bibitem{SLAC}  
L.~W.~Whitlow {\it et al.},
Phys. Lett. B {\bf 282}, 475 (1992).

\bibitem{NMCp}
M.~Arneodo {\it et al.},
Nucl. Phys. B {\bf 483}, 3 (1997).

\bibitem{NMCdop}
M.~Arneodo {\it et al.},
Nucl. Phys. B {\bf 487}, 3 (1997).
           
\bibitem{HERA2}
H.~Abramowicz {\it et al.},
Eur. Phys. J. C {\bf 75}, 580 (2015).

\bibitem{HERMES} 
A.~Airapetian {\it et al.},
J. High Energy Phys. {\bf 05} (2011) 126.

\bibitem{Malace}
S.~P.~Malace {\it et al.},
Phys. Rev. C {\bf 80}, 035207 (2009).

\bibitem{CDF_e}
D.~Acosta {\it et al.},
Phys. Rev. D {\bf 71}, 051104(R) (2005).

\bibitem{CDF_W}
T.~Aaltonen {\it et al.},
Phys. Rev. Lett. {\bf 102}, 181801 (2009).

\bibitem{CDFZ}
T.~Aaltonen {\it et al.}, 
Phys. Lett. B {\bf 692}, 232 (2010).

\bibitem{D0Z}
V.~M.~Abazov {\it et al.},
Phys. Rev. D {\bf 76}, 012003 (2007).

\bibitem{CDFjet2}
T.~Aaltonen {\it et al.},
Phys. Rev. D {\bf 78}, 052006 (2008).

\bibitem{D0jet2}
V.~M.~Abazov {\it et al.},
Phys. Rev. Lett. {\bf 101}, 062001 (2008);
%
B.~Abbott {\it et al.},
Phys. Rev. Lett. {\bf 86}, 1707 (2001).

\bibitem{D0gamjet}
V.~M.~Abazov {\it et al.},
Phys. Lett. B {\bf 666}, 435 (2008).

\bibitem{Alekhin06} 
S.~Alekhin, K.~Melnikov and F.~Petriello,
Phys. Rev. D {\bf 74}, 054033 (2006).

\bibitem{E605}
G.~Moreno {\it et al.},  
Phys. Rev. D {\bf 43}, 2815 (1991).

\bibitem{PedroIC15}
P.~Jimenez-Delgado, T.~J.~Hobbs, J.~T.~Londergan and W.~Melnitchouk,
Phys. Rev. Lett. {\bf 114}, 082002 (2015).

\bibitem{PDG}
K.~A.~Olive {\it et al.},
Chin. Phys. C {\bf 38}, 090001 (2014).

\bibitem{ABM11}
S.~Alekhin, J.~Bl\"umlein and S.-O.~Moch,
Phys. Rev. D {\bf 86}, 054009 (2012).

\bibitem{HMMT16}
L.~A.~Harland-Lang, A.~D.~Martin, P.~Motylinski and R.~S.~Thorne,
arXiv:1601.03413 [hep-ph].

\bibitem{MARATHON}
Jefferson Lab Experiment C12-10-103 [MARATHON],
G.~G.~Petratos, J.~Gomez, R.~J.~Holt and R.~D.~Ransome,
spokespersons.
 
\bibitem{BONUS12}
Jefferson Lab Experiment E12-10-102 [BONUS12],
S.~B\"ultmann, M.~E.~Christy, H.~Fenker, K.~Griffioen, C.~E.~Keppel,
S.~Kuhn and W.~Melnitchouk,
spokespersons.
 
\bibitem{SOLID}
Jefferson Lab Experiment E12-10-007 [SoLID],
P.~Souder, spokesperson.

\bibitem{Accardi16}
A.~Accardi, 
arXiv:1602.02035 [hep-ph]

\bibitem{MRST03}
A.~D.~Martin, R.~G.~Roberts, W.~J.~Stirling and R.~S.~Thorne,
Eur. Phys. J. C {\bf 35}, 325 (2004).

\bibitem{ABKM09} 
S.~Alekhin, J.~Bl\"umlein, S.~Klein and S.-O.~Moch,
Phys. Rev. D {\bf 81}, 014032 (2010).

\bibitem{Brady12} 
L.~T.~Brady, A.~Accardi, W.~Melnitchouk and J.~F.~Owens,
JHEP {\bf 1206}, 019 (2012).

\bibitem{PDFLHC16}
A.~Accardi {\it et al.},
arXiv:1603.08906 [hep-ph].

\bibitem{Afnan00}
I.~R.~Afnan {\it et al.},
Phys. Lett. B {\bf 493}, 36 (2000);
Phys. Rev. C {\bf 68}, 035201 (2003).

\bibitem{Pace01}
E.~Pace, G.~Salme, S.~Scopetta and A.~Kievsky,
Phys. Rev. C {\bf 64}, 055203 (2001).

\bibitem{SSS02}
M.~M.~Sargsian, S.~Simula and M.~I.~Strikman,
Phys. Rev. C {\bf 66}, 024001 (2002).

\end{thebibliography}
\end{document}